\documentclass[preprint,5p]{elsarticle}

\usepackage{amsmath}
\usepackage{amsfonts}
\usepackage{amssymb}
\usepackage{physics}
\usepackage{slashed}
\usepackage{simplewick}

\usepackage{graphicx}
\usepackage{float}
\usepackage{subcaption}

\usepackage{listings}

\usepackage{tikz}
\usetikzlibrary{calc}
\usetikzlibrary{patterns}
\usetikzlibrary{patterns}
\usetikzlibrary{positioning,decorations.pathmorphing,decorations.markings}

\usepackage{soul}
\usepackage[utf8]{inputenc}

\usepackage{feynmp-auto}
\usepackage{multirow}

\newcommand{\p}{\ensuremath{\prime}}

\usepackage{widetext}
\usepackage{xcolor}

\usepackage{graphicx}
\usepackage{float}
\usepackage{subcaption}
\usepackage{overpic}

\usepackage{hyperref}

\journal{Physics Letters B}

\begin{document}
\begin{fmffile}{feynDiags}

\begin{frontmatter}

\title{Model Dependence of the Pion Form Factor Extracted from Pion Electro-production}

\author{Robert J. Perry}
\author{Ay{\c s}e K{\i}z{\i}lers{\" u} }
\author{Anthony W. Thomas}
\address{CSSM and ARC Centre of Excellence for Particle Physics at the Tera-scale, Department of Physics, University of Adelaide, Adelaide SA 5005, Australia}

\begin{abstract}
In 2008 the Jefferson Laboratory $F_\pi$ Collaboration released results for the pion electromagnetic form factor, which they extracted from pion electro-production data. The measured values for the pion form factor are model dependent, and require the use of the Vanderhaeghen, Guidal and Laget Regge Model for their extraction. While agreement between this model and data is impressive, the theoretical implementation of gauge invariance is less satisfying. We would like to establish how well the extracted form factor corresponds to the true form factor. To do this, we use a simple toy model, which imposes gauge invariance in a more theoretically satisfying way. The model form factor is extracted from our model cross section using the method employed by the $F_\pi$ Collaboration to extract the experimental pion form factor. We conclude that the reconstructed model form factor is a reasonable representation of the true model form factor for the kinematics chosen, although we note that the extracted form factor is smaller than the true form factor. This suggests that current extracted values of pion form factor may be overestimated.
\end{abstract}
%

%\begin{keyword}
%chiral symmetry \sep electromagnetic form factors \sep NJL Model \sep pion corrections 
%\PACS 
%13.40.Em \sep % Electric and magnetic moments 
%13.40.Gp \sep % electromagnetic form factors 
%14.20.Jn \sep % Hyperons
%25.30.Bf \sep % Elastic electron scattering
%12.39.-x %  	Phenomenological quark models
%\end{keyword}
%
\end{frontmatter}

\section{Introduction}

%\begin{itemize}
%\item What is the point of the paper?
%\begin{itemize}
%\item the experimental extraction of the pion form factor
%\end{itemize}
%\item It is important because the pion is the pseudo goldstone boson of QCD; characteristic light mass signals the dynamical breaking of chiral symmetry.
%\item Characteristic lightness leads to it being the longest contribution to the $NN$ potential.
%\item In order to understand low to intermediate energy QCD, we must understand the pion.
%\item We may probe the elastic structure of the pion using an electromagnetic probe. 
%\begin{itemize}
%\item At low photon virtuality, we have $\chi$PT, LQCD. Predicts charge radii etc.
%\item At high photon virtuality ($Q^2>>\Lambda_\text{QCD}$), we have PQCD.
%\item In intermediate region, ??? We don't have many good approaches.
%\end{itemize}
%\end{itemize}

The pion is the pseudo Goldstone Boson associated with the dynamical breaking of the approximate chiral symmetry of QCD. Because of its characteristically light mass, the pion yields the longest range contribution to low energy hadronic observables. This dominance at low energies means both that the pion is the most easily studied meson and also that a deep understanding of the pion is necessary to investigate the low energy, non-perturbative behavior of QCD.

One clear window into the complicated behaviour of the pion is its electromagnetic form factor, $F_\pi(Q^2)$. This Lorentz invariant structure function encodes the non-pertur-  
bative behaviour of the electromagnetically charged partons inside the pion and may be related to the the pion's transverse charge distribution~\cite{Miller:2010nz}. Many different complementary theoretical descriptions exist of the pion form factor. At low photon virtuality, the pion form factor may be calculated from first principles using Lattice QCD but as the photon virtuality increases the extraction of the pion form factor becomes more difficult. Novel techniques have allowed extraction of the pion form factor out to about 6 GeV$^2$~\cite{Chambers:2017tuf}. Models based on QCD may be extended to higher momenta. For example, the pion electromagnetic form factor was reasonably well predicted in the NJL Model~\cite{Cloet:2014rja}. More complicated models, based on the Dyson-Schwinger approch are also able to predict to good accuracy the current experimental data~\cite{Maris:2000sk}. Finally, in the large momentum limit, the result of Lepage and Brodsky~\cite{Lepage:1979zb} is expected to hold~:
\begin{equation}
Q^2F_\pi(Q^2)\to 16\pi f_\pi^2\alpha_s(Q^2),
\end{equation}
where $f_\pi\approx0.132$ GeV is the pion decay constant, and $\alpha_S(Q^2)$ is the strong coupling constant. Since this result is derived in the context of perturbative QCD, probing the high momentum behaviour of $F_\pi$ gives information on the transition from the non-perturbative regime to the perturbative regime in QCD.
Agreement with this relation would give confidence that the asymptotic limit of QCD is understood. As is shown below in Fig.~\ref{fig:1}, the current data indicates that we have not yet probed high enough energies for the above relation to hold.

%
%\begin{itemize}
%\item Theoretical limits on infinite momentum limit of pion form factor.
%\item Dispersion Theory.
%\end{itemize}
The pion form factor is also an experimentally difficult observable to measure at larger momenta. At low photon virtuality, the pion form factor may be measured directly from elastic $e^-+\pi^+$ scattering. However, due to kinematic limitations of the pion beam, this approach only allows an extraction of the pion form factor up to approximately 0.3 GeV$^2$~\cite{Amendolia:1986wj}. Thus for larger momentum transfer values, another technique must be used. We call this region between the failure of direct measurement in $e^-+\pi^+$ scattering and the onset of perturbative QCD the \textit{intermediate} momentum region (see Fig.~\ref{fig:1}). Modern extractions of the pion electromagnetic form factor in this intermediate region utilize pion electro-production. 

\begin{figure}
\centering
\includegraphics[scale=0.4]{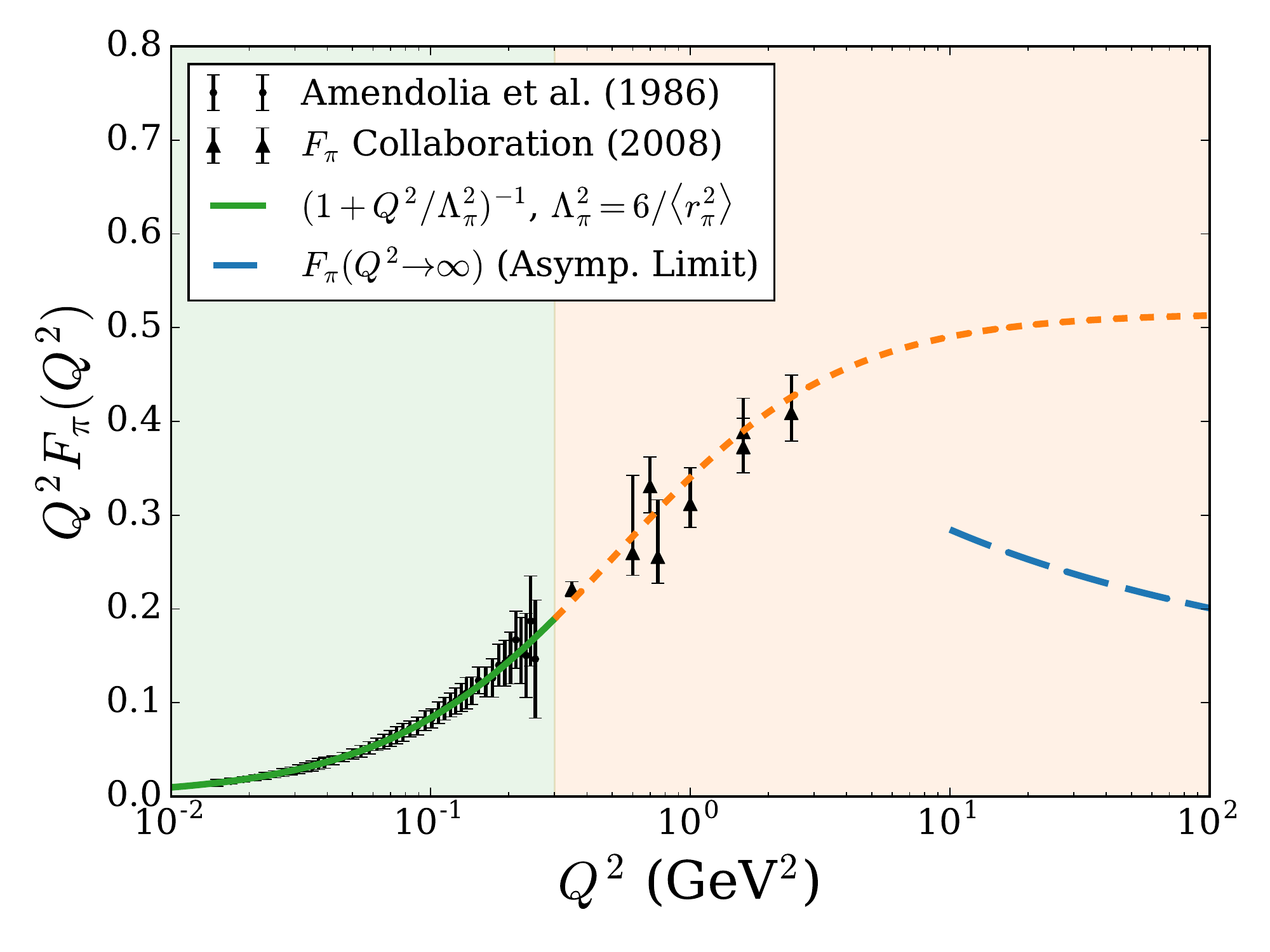}
\caption{(Colour online) Experimental data for the pion electromagnetic form factor. Low energy data (shown in the green region) is taken from Ref.~\cite{Amendolia:1986wj}, and intermediate energy data (shown in the orange region) is taken from Ref.~\cite{Huber:2008id}. Note that the intermediate momentum region is defined to be the region between the failure of direct measurement of $e^-\pi^+$ elastic scattering and the scale at which perturbative QCD becomes applicable. For comparison, the asymptotic limit of the electromagnetic form factor is also shown. The energy scale at which this result becomes applicable is still an open question.}\label{fig:1}
\end{figure}

In pion electro-production, information about the pion form factor is obtained by scattering an electron off the pion cloud of the nucleon. There are a number of complications to this approach which must be addressed in order to extract the pion form factor from this process.

Firstly, there is an interference between the $t$-channel term, which contains the pion form factor, and the $s$- and $u$-channel terms, which contain the nucleon form factors. Thus any extraction of the pion form factor from this process must understand how these interference terms effect the overall measured cross section.

Secondly, the pion is initially, off-shell. The pion's virtuality is measured by the $t$ Mandelstam variable. Importantly, due to the kinematics of pion electro-production, $t$ is kinematically constrained to be negative, whereas the on-shell pion form factor should be obtained by probing an on-shell pion, that is, at $t=m_\pi^2>0$. Thus, while it is clear that this process should be able to give us information about the pion form factor, it is not immediately obvious how closely related the off-shell pion is to the on-shell pion measured in direct $e^-+\pi^+$ scattering. Previously, this question has been addressed in the context of a Bethe-Salpeter approach~\cite{Qin:2017lcd}. There, the authors found that the `pion form factor' increased in magnitude as the pion deviated further off-shell. We place pion form factor in quotes here to emphasise, as do the authors of the paper, that one may only truly talk about the pion form factor when the pion is on-shell, since there is no unique definition of an off-shell state. This result would seem to indicate that care must be taken when extracting the pion form factor to ensure that variation of the pion form factor due to pion's `off-shellness' is minimized.

\begin{figure}
\includegraphics[scale=0.4]{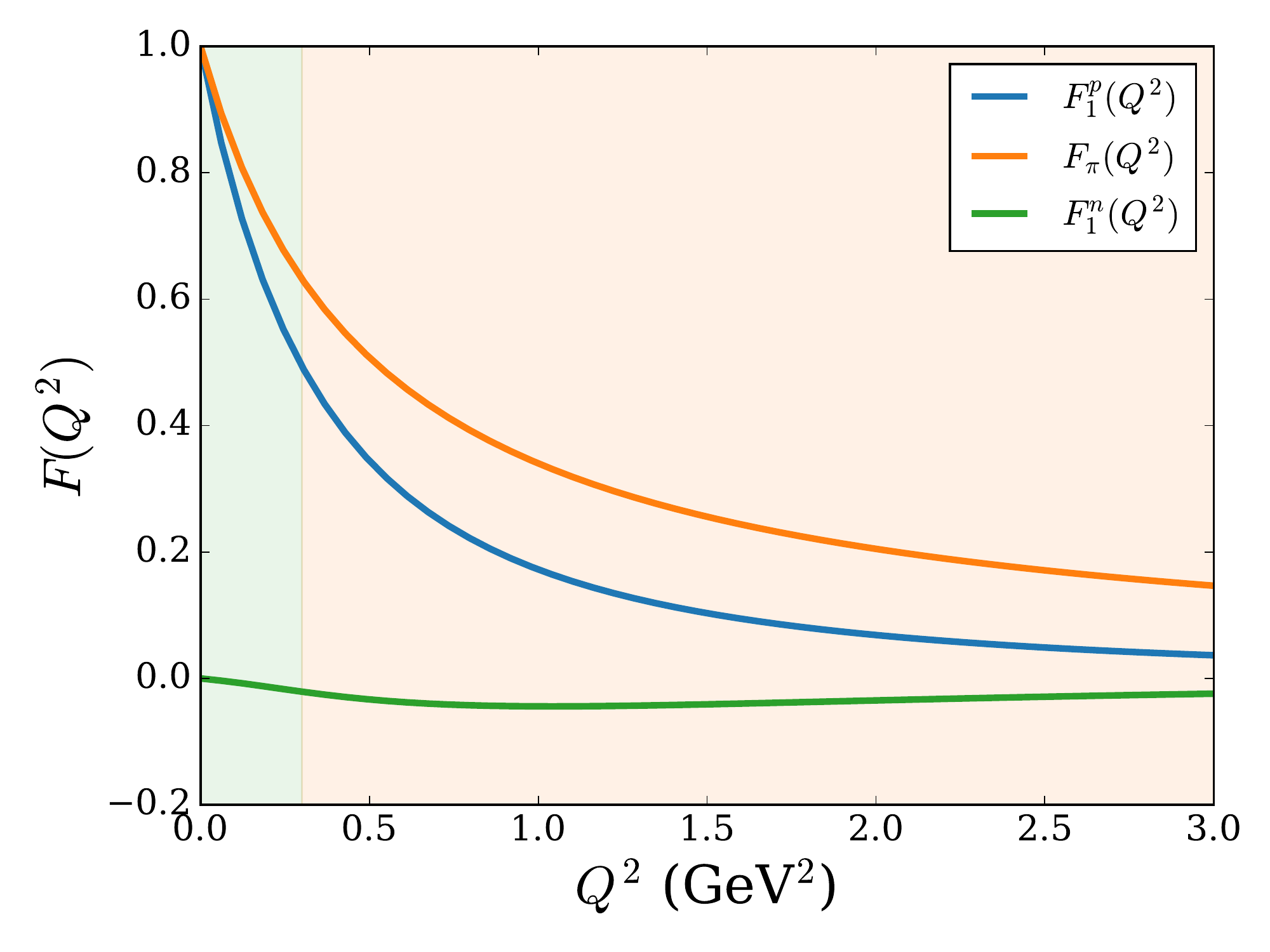}
\caption{(Colour online) Comparison of the pion electromagnetic form factor with the proton and neutron Dirac form factors. As is conventional, we have parameterised the pion form factor with a monopole form $F_\pi(Q^2)=(1+Q^2/\Lambda_\pi^2)^{-1}$, while we parameterise the proton Dirac form factor with a dipole form $F_1^p(Q^2)=(1+Q^2/\Lambda_p^2)^{-2}$. For the neutron, we use the Galster parameterization~\cite{Galster:1971kv}. The parameters $\Lambda_\pi$ and $\Lambda_p$ may be related to the electromagnetic charge radius of the respective particle. The shaded regions correspond to the same momentum regions shown in Fig.~\ref{fig:1}.}\label{fig:2}
\end{figure}

In order to extract the pion form factor from the electro-production data, a model of the differential cross section must be used. The $F_\pi$ Collaboration use the Vanderhaeghen, Guidal and Laget (VGL ) Model, a Regge Model in which the pole-like propagators of a Born Term Model are replaced by Regge propagators where the single particles are replaced by the exchange of a family of particles with the same internal quantum numbers.   In order to incorporate the extended structure of the pion in the VGL Model, the pion's electromagnetic form factor is included in the matrix element, which is given by
\begin{equation}
i\mathcal{M}_{\text{VGL}}^\mu=i\mathcal{M}_{\text{R}}^\mu F_\pi(Q^2),
\end{equation}
where $i\mathcal{M}_{\text{R}}^\mu$ is the Reggized gauge invariant amplitude for the scattering of point-like nucleons and pions, and $F_\pi(Q^2)$ is the electromagnetic form factor. A second term describes the exchange of a rho meson instead of the pion in the $t$-channel, but it is irrelevant for the point we are trying to make, so we ignore it here. This matrix element leads to a differential cross section of the form
\begin{equation}
\left(\frac{d\sigma_L}{dt}\right)_\text{VGL}=\big(F_\pi(Q^2)\big)^2\, \left(\frac{d\sigma_L}{dt}\right)_\text{R},
\end{equation}
where $\left(d\sigma_L/dt\right)_R$ is termed the \textit{longitudinal} cross section, obtained from the matrix element $i\mathcal{M}_\text{R}^\mu$. In principle, only the $t$-channel diagram can contribute to the form factor $F_\pi(Q^2)$, since the $s$- and 
$u$-channel diagrams give information about the nucleon form factors. To write the matrix element as is done above, one must approximate the pion and proton Dirac form factors to be equal. For a comparison of the relevant pion and proton form factors in the region probed by pion electro-production, see Fig.~\ref{fig:2}.

%\begin{itemize}
%\item Historically, the form factor has been incorporated at the level of cross sections ($d\sigma/dt=|F|^2d\sigma/dt_\text{p.l}$), rather than at amplitudes.
%\item I think that the $t$-channel is the only diagram which survives to contribute to the Longitudinally polarized cross section.
%\begin{itemize}
%\item Conserve Helicity?
%\end{itemize} 
%\item This approach is clearly wrong when the form factor is introduced at the level of the amplitude.
%\item The $F_\pi$ in the VGL Model is an effective $F_\pi$, which parameterizes \textit{all} the structure not described explicitly by the point-like cross section. 
%\end{itemize}

Following the above discussion, we now  wish to test whether this approximation leads to inconsistencies in the extracted pion form factor. To do this, we choose a simple toy model, which allows us to calculate form factors and cross section exactly. We then use the method employed by the $F_\pi$ Collaboration to attempt to extract the toy model's pion form factor from our cross section. This allows us to see how well we are able to extract the form factor using this approach. While the use of a toy model prevents us from making direct statements about the physical extracted pion form factor, we suggest that the conclusions drawn from our toy model may carry over qualitatively to the physical form factor.

%\begin{itemize}
%\item Summary of importance of pion:
%\begin{itemize}
%\item Pseudo-goldstone boson
%\item Long time known as important for low energy observables: chiral perturbation theory.
%\item Important to understand the structure: Elastic structure encoded in electromagnetic form factors.
%\item Theoretical and experimental effort to understand the form factor
%\item LQCD, SDE, Quark Models, 
%\item Ongoing study: J-Lab doing experiment at higher $Q^2$ to measure the form factor here.
%\end{itemize}
%\item J-Lab published results utilize the VGL Model:
%\begin{itemize}
%\item EFT description
%\item Born Term Model+Reggised trajectory
%\item Form factor multiplies amplitude.
%\item Leads to good fit to photoproduction and electroproduction over a range of $Q^2$ and $t$.
%\item \textbf{but!} Implementation of EM gauge invariance unnatural.
%\begin{itemize}
%\item Two questions: 1) Can we find a more natural way to express the gauge invariance? and 2) How confident can we be in the current extraction?
%\end{itemize} 
%\end{itemize}
%\item 
%\end{itemize}

\section{Kinematics and Preliminaries}

\begin{figure}[H]
\centering
\vspace{10pt}
\begin{fmfgraph*}(60,40)
\fmfleft{l1,l2}
\fmfright{r1,r2}
\fmf{fermion,width=2,label.side=right}{l1,v1}
\fmf{fermion,width=2}{v1,r1}
\fmf{photon}{v1,l2}
\fmf{dashes,label.side=left}{v1,r2}
\fmfv{decoration.shape=circle,decoration.filled=shaded,decoration.size=30,label.dist=15}{v1}
\fmflabel{$q$}{l2}
\fmflabel{$p$}{l1}
\fmflabel{$p^\p$}{r1}
\fmflabel{$p_\pi$}{r2}
\end{fmfgraph*}
\begin{tikzpicture}
\node at (0,0) {};
\node[align=center] at (1.5,0.5) {time};
\draw[->] (1,0) -- (2,0);
\end{tikzpicture}
\vspace{10pt}
\caption{Hadronic component on pion electro-production in the one-photon exchange approximation (o.p.e.a.). Overall conservation of momentum gives $p+q=p^\p+p_\pi$. This sets the direction of the external momenta.}\label{fig:momDir}
\end{figure}
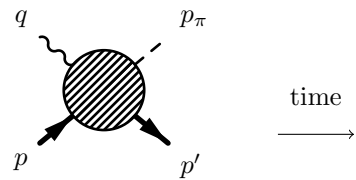

Before discussing the VGL Model in more detail, we first introduce our conventions for kinematic variables and structure functions. We label external momenta as shown in Fig.~\ref{fig:momDir}, where overall momentum conservation gives \\
$p+q=p^\p+p_\pi$. This convention for external particle momenta determines how momentum flows into the loop diagrams. 

We define the Mandelstam variables
\begin{align}
s&=p_s^2 = (p+q)^2    =(p^\p+p_\pi)^2 \equiv W^2,
\\
t&=p_t^2  = (p_\pi-q)^2=(p-p^\p)^2,
\\
u&=p_u^2 = (p-p_\pi)^2=(p^\p-q)^2.
\end{align}
Finally, we define $Q^2=-q^2$ so that spacelike momenta are positive. These three momenta ($Q^2$, $W$ and $t$) allow one to fully describe the cross section. The measured unpolarized differential cross section may be separated according to the polarization states of the virtual photon into transverse ($T$), longitudinal ($L$) polarizations, as well as two interference terms ($LT$ and $TT$)~\cite{Blok:2008jy}:
\begin{equation}
\begin{split}
(2\pi)\frac{d^2\sigma}{dtd\phi}\,\,=\,\,&\frac{d\sigma_T}{dt}+\epsilon\frac{d\sigma_L}{dt}  
\\
+&\sqrt{2\epsilon(\epsilon+1)}\frac{d\sigma_{LT}}{dt}\cos\phi+\epsilon\frac{d\sigma_{TT}}{dt}\cos2\phi  ,
\end{split}
\end{equation}
where $\epsilon$ is a measure of the virtual photon polarization, and is related to experimental quantities via
\begin{equation}
\epsilon=\bigg(1+\frac{2|\vec{q}|^2}{Q^2}\tan^2\frac{\theta_e}{2}\bigg)^{-1},
\end{equation}
$\vec{q}$ is the three-momentum of the virtual photon, and $\theta_e$ is the angle between the initial and final electron three-momentum. This decomposition is important because it is well known that the $t$-channel pion exchange diagram dominates the longitudinal differential cross section $d\sigma_L/dt$~\cite{Kaskulov:2008xc}. It has been shown previously that rho meson exchange is suppressed in the longitudinal cross section by approximately an order of magnitude (see Fig.~4, Ref.~\cite{Vanderhaeghen:1997ts}). Thus a model which precisely predicts the longitudinal differential cross section has a good chance of extracting the pion form factor.

\section{The VGL Model}

%\begin{itemize}
%\item FIX: Inclusion of rho terms.
%\begin{itemize}
%\item Not really relevant for us as it doesn't effect longitudinal cross section.
%\item Still mention it!
%\end{itemize}
%\end{itemize}

%Since we are only interested in improving the implementation of gauge invariance, we will deal with the Born Term version of the pion propagator, rather than the Reggized version.

Before proposing alternative approaches, we must first understand the VGL Model. Originally developed by Vanderhaeghen Guidal and Laget as a model of pion photo-production~\cite{Guidal:1997by}, it was quickly realized that the generalization to electro-production was straightforward, leading to the so-called VGL Model~\cite{Vanderhaeghen:1997ts}. The VGL Model is based on the $t$-channel pion exchange Born Diagram, shown in Fig.~\ref{fig:4}. The 
$t$-channel diagram is not gauge invariant on its own, and requires the inclusion of the $s$-channel and Kroll-Ruderman terms to restore gauge invariance. The $t$-channel diagram in which a rho meson is exchanged instead of a pion is also included. This diagram is independently gauge invariant, so no accompanying diagrams must be added to preserve gauge invariance. In order to improve agreement between the model and data, the pion propagator is \emph{Reggeized}, which amounts to replacing the pion and rho meson propagators $S_F^{\pi/\rho}(t)$ with its Reggeized version $S_\text{R}^{\pi/\rho}(t)$. In order to understand the VGL Model, it helps to begin by examining the Born Term Model upon which it is based.

\subsection{The Born Tern Model}

Using the Feynman Rules outlined in Ref.~\cite{Ji:2013bca}, we can show that the Born Term Model (BTM) arising from this Lagrangian is given by
\begin{equation}
i\mathcal{M}_\text{BTM}^\mu=i\mathcal{M}_{s}^\mu+i\mathcal{M}_{t}^\mu+i\mathcal{M}_{\text{K.R.}}^\mu,
\end{equation}
where the associated diagrams are given in Fig.~\ref{fig:4}. We have ignored the rho meson term shown above, as it is gauge invariant on it's own, and adds nothing to the understanding of the VGL Model. The three terms are denoted $i\mathcal{M}_{s}^\mu$, $i\mathcal{M}_{t}^\mu$ and $i\mathcal{M}_{\text{K.R.}}^\mu$, respectively:
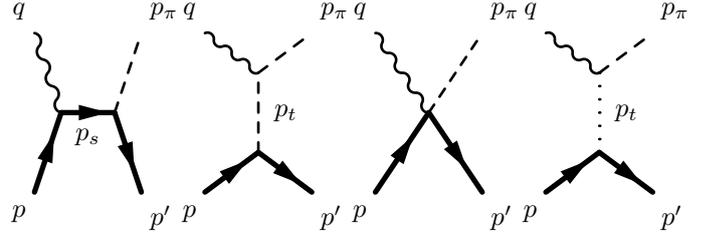
\begin{figure}
\centering
\begin{subfigure}{0.24\columnwidth}
\begin{fmfgraph*}(50,60)
\fmfleft{l1,l2}
\fmfright{r1,r2}
\fmf{fermion,width=2}{l1,v1}
\fmf{fermion,width=2,label=$p_s$}{v1,v2}
\fmf{fermion,width=2}{v2,r1}
\fmf{dashes}{v2,r2}
\fmf{photon,label.side=right}{v1,l2}
\fmflabel{$q$}{l2}
\fmflabel{$p^\p$}{r1}
\fmflabel{$p$}{l1}
\fmflabel{$p_\pi$}{r2}
%\fmfdot{v1,v2}
\end{fmfgraph*}
\end{subfigure}
\begin{subfigure}{0.24\columnwidth}
\begin{fmfgraph*}(50,60)
\fmfleft{l1,l2}
\fmfright{r1,r2}
\fmf{fermion,width=2}{l1,v1,r1}
\fmf{dashes,label=$p_t$,label.side=right}{v1,v2}
\fmf{dashes}{v2,r2}
\fmf{photon,label.side=right}{v2,l2}
\fmflabel{$q$}{l2}
\fmflabel{$p^\p$}{r1}
\fmflabel{$p$}{l1}
\fmflabel{$p_\pi$}{r2}
%\fmfdot{v1,v2}
\end{fmfgraph*}
\end{subfigure}
\begin{subfigure}{0.24\columnwidth}
\begin{fmfgraph*}(50,60)
\fmfleft{l1,l2}
\fmfright{r1,r2}
\fmf{fermion,width=2}{l1,v2,r1}
\fmf{dashes}{v2,r2}
\fmf{photon}{v2,l2}
\fmflabel{$q$}{l2}
\fmflabel{$p^\p$}{r1}
\fmflabel{$p$}{l1}
\fmflabel{$p_\pi$}{r2}
%\fmfdot{v2}
\end{fmfgraph*}
\end{subfigure}
\begin{subfigure}{0.24\columnwidth}
\begin{fmfgraph*}(50,60)
\fmfleft{l1,l2}
\fmfright{r1,r2}
\fmf{fermion,width=2}{l1,v1,r1}
\fmf{dots,label=$p_t$,label.side=right}{v1,v2}
\fmf{dashes}{v2,r2}
\fmf{photon,label.side=right}{v2,l2}
\fmflabel{$q$}{l2}
\fmflabel{$p^\p$}{r1}
\fmflabel{$p$}{l1}
\fmflabel{$p_\pi$}{r2}
%\fmfdot{v1,v2}
\end{fmfgraph*}
\end{subfigure}
\vspace{10pt}
\caption{Born Term Model for pion electro-production. The pion form factor is measured in pion electroproduction via the $t$-channel diagram. The second $t$-channel diagram corresponds to the exchange of a virtual rho meson. There is no $u$ channel diagram because in our effective field theory, the neutron is neutral at tree level.}\label{fig:4}
\end{figure}
\begin{align}
i\mathcal{M}_{s}^\mu=&\frac{g_A}{\sqrt{2}f_\pi}\overline{u}_N(p^\p,s^\p)\gamma_5\slashed{p}_\pi S_F^N(p_s)
\\
&\times(-ie\gamma^\mu)u_N(p,s),
\\
i\mathcal{M}_{t}^\mu=&\frac{g_A}{\sqrt{2}f_\pi}\overline{u}_N(p^\p,s^\p)\gamma_5\slashed{p}_tu_N(p,s)S_F^\pi(p_t)
\\
&\times(-ie)(p_t+p_\pi)^\mu,
\\
i\mathcal{M}_{\text{K.R.}}^\mu=&-\frac{g_Ae}{\sqrt{2}f_\pi}\overline{u}_N(p^\p,s^\p)\gamma_5\gamma^\mu u_N(p,s).
\end{align}
Thus the Born Term Model matrix element is
\begin{equation}\label{eq:BTM}
\begin{split}
i\mathcal{M}_{\text{BTM}}^\mu=&\frac{g_A}{\sqrt{2}f_\pi}\overline{u}_N(p^\p,s^\p)\gamma_5\slashed{p}_\pi S_F^N(p_s)(-ie\gamma^\mu)u_N(p,s)
\\
&+\frac{g_A}{\sqrt{2}f_\pi}\overline{u}_N(p^\p,s^\p)\gamma_5\slashed{p}_tu_N(p,s)S_F^\pi(p_t)
\\
&\times (-ie)(p_t+p_\pi)^\mu
\\
&-\frac{g_Ae}{\sqrt{2}f_\pi}\overline{u}_N(p^\p,s^\p)\gamma_5\gamma^\mu u_N(p,s).
\end{split}
\end{equation}

\subsection{Transforming the Born Term Model to the VGL Model}

One may obtain the VGL Model by first Reggeizing this amplitude, and then further multiplying the amplitude by the pion form factor. Importantly, in replacing the Feynman propagators with their Regge versions, gauge invariance must be preserved. One may understand the Reggeization of the amplitude used in the VGL Model as the multiplication of the Born Term Model amplitude by a \textit{single momentum dependent factor} $S_F^{\pi-1}(p_t)S_R^\pi(p_t)$, where $S_F^{\pi}(p_t)$ is the Feynman propagator, and $S_R^\pi(p_t)$ is the Reggeized pion proagator. Thus
\begin{equation}\label{eq:regge}
i\mathcal{M}_{\text{R}}^\mu=S_F^{\pi-1}(p_t)S_R^\pi(p_t)\big[i\mathcal{M}_{\text{BTM}}^\mu\big].
\end{equation}
More will be said about this procedure later. In particular, we will explain why it is helpful to think of the Reggeization step a multiplicative procedure on the amplitude. Reggeizing the amplitude in this way leads to the Reggeized matrix element $i\mathcal{M}_{\text{R}}^\mu$:
\begin{equation}
\begin{split}
i\mathcal{M}_{\text{R}}^\mu=&\frac{g_A}{\sqrt{2}f_\pi}\overline{u}_N(p^\p,s^\p)\gamma_5\slashed{p}_\pi S_F^N(p_s)S_F^{\pi-1}(p_t)S_R^\pi(p_t)
\\
&\times (-ie\gamma^\mu)u_N(p,s)
\\
&+\frac{g_A}{\sqrt{2}f_\pi}\overline{u}_N(p^\p,s^\p)\gamma_5\slashed{p}_tu_N(p,s)S_R^\pi(p_t)
\\
&\times (-ie)(p_t+p_\pi)^\mu
\\
&-\frac{g_Ae}{\sqrt{2}f_\pi}\overline{u}_N(p^\p,s^\p)S_F^{\pi-1}(p_t)S_R^\pi(p_t)\gamma_5\gamma^\mu u_N(p,s).
\end{split}
\end{equation}

The pion form factor is then introduced as
\begin{equation}
i\mathcal{M}_{\text{VGL}}^\mu=i\mathcal{M}_{\text{R}}^\mu F_\pi(Q^2),
\end{equation}
where $F_\pi(Q^2)$ is the electromagnetic form factor. Having briefly discussed the VGL Model, we will now explain the process by which the pion form factor is extracted from experimental data.

\subsection{Fitting the VGL Model to Data}\label{sec:fittingVGL}

We can now summarize the procedure used by the $F_\pi$ Collaboration to fit the VGL Model to experimental data. The functional form of the pion form factor is taken to be the monopole form:
\begin{equation}
F_\pi(Q^2)=\frac{1}{1+Q^2/\Lambda_\pi^2},
\end{equation}
and the transition form factor for the $\rho$ is assumed to have the same functional form:
\begin{equation}
F_{\gamma\rho\pi}(Q^2)=\frac{1}{1+Q^2/\Lambda_\rho^2},
\end{equation}
where $\Lambda_\pi^2$ and $\Lambda_\rho^2$ are the only free parameters in the model. As mentioned previously, the longitudinal cross section is insensitive to the rho meson, so effectively only $\Lambda_\pi^2$ must be fit to obtain the longitudinal cross section.

\begin{figure}[H]
\centering
\includegraphics[scale=0.4]{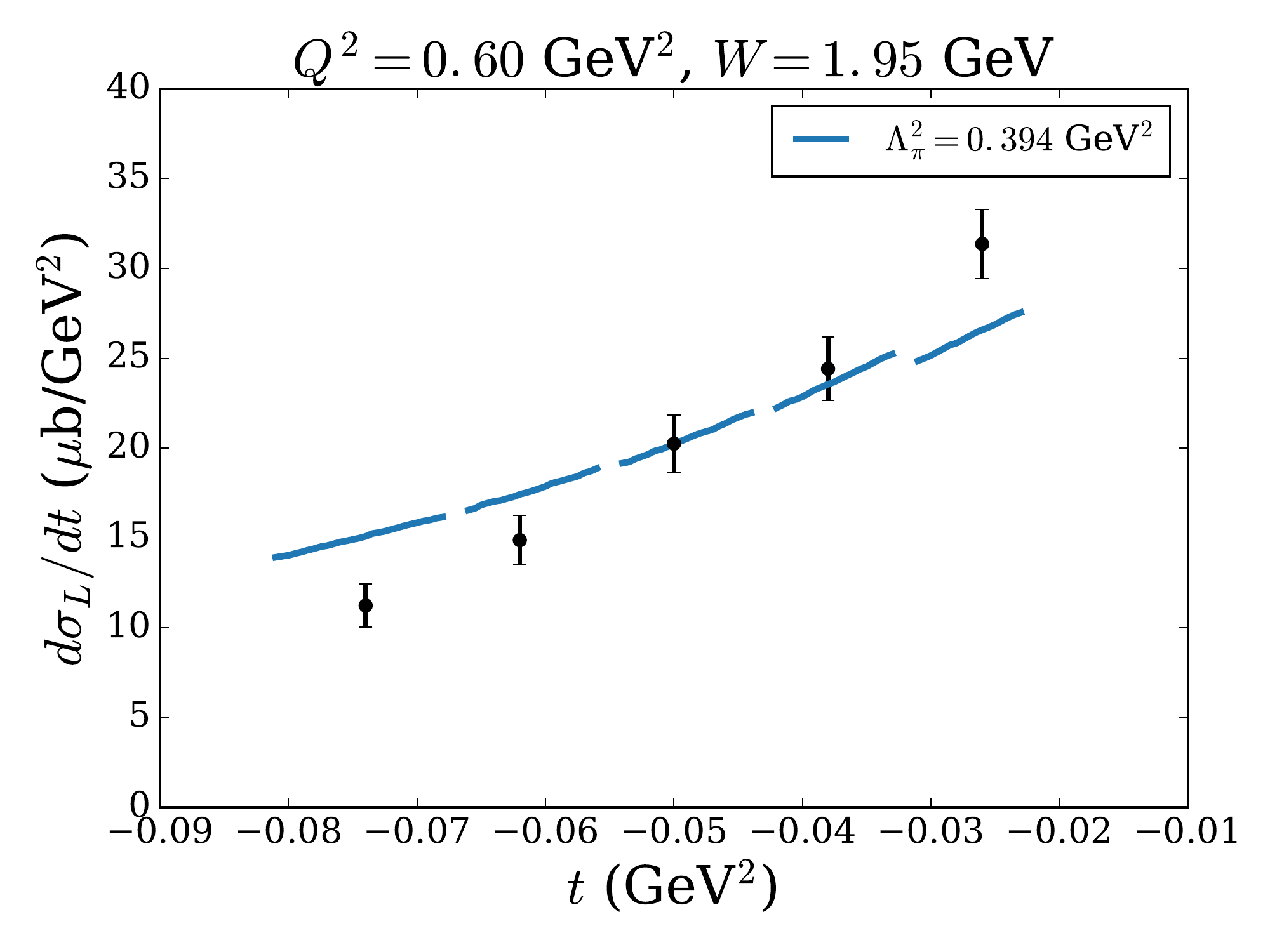}
\includegraphics[scale=0.4]{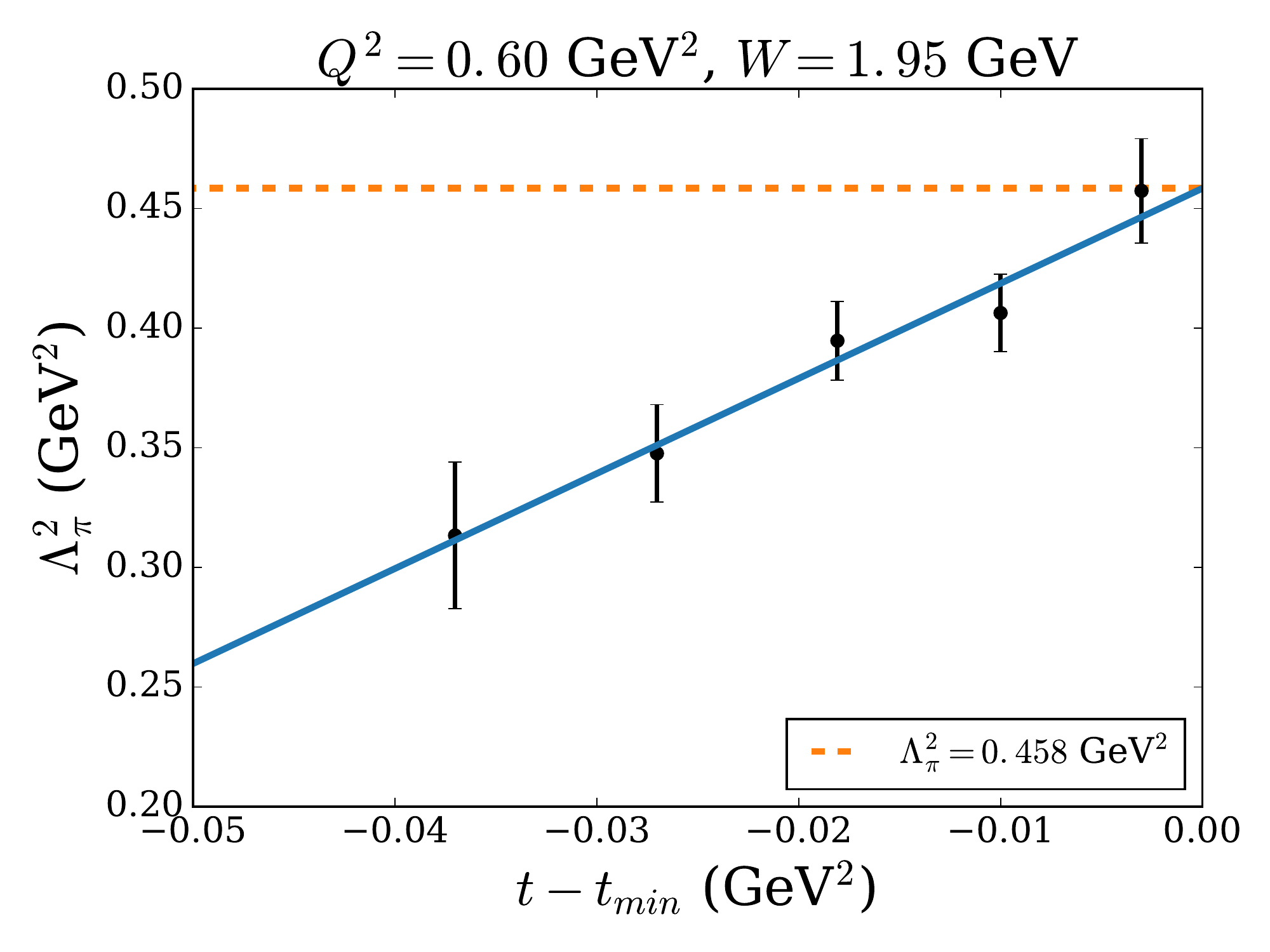}
\caption{Plots adapted from the $F_\pi$ Collaboration's extraction of the pion form factor (Ref.~\cite{Huber:2008id}, Fig.~2, Fig.~5). The top plot shows the fitted longitudinal cross section, compared to experimental data. Note that the theory curve for the longitudinal cross section is plotted for a single value of $\Lambda_\pi^2$ to demonstrate the general agreement of the VGL Model with data. As explained, when performing the extraction, the model is fit to each data point independently. The bottom plot shows the corresponding extracted values of $\Lambda_\pi^2$. The best fit value for $\Lambda_\pi^2$ for this set of kinematics is $\Lambda_\pi^2=0.458\pm0.031_{-0.068}^{+0.255}$ GeV$^2$~\cite{Huber:2008id}.}\label{fig:extraction}
\end{figure}

The differential cross section is first measured at a range of $Q^2$ and $W$ values for small $|t|$, and then each longitudinal cross section data point is fit independently to the VGL Model. Thus for each data point, there is a corresponding extracted $\Lambda_\pi^2$. This is shown in the second plot of Fig.~\ref{fig:extraction}. Note that in general, data points measured at smaller values of $t$ tend to result in  larger values of $\Lambda_\pi^2$, and thus larger values of $F_\pi(Q^2)$. It has been suggested that this is due to interfering backgrounds not included in the VGL model~\cite{Huber:2008id}. In practice, an extrapolation of $\Lambda_\pi^2$ to the minimum allowed $t$ value is performed and it is this value of $\Lambda_\pi^2$ which is taken to correspond to the best estimate of $F_\pi(Q^2)$. These values are shown in Fig.~\ref{fig:1}.

While the agreement between the VGL Model and data is quite good, as we have shown, there is room to improve the implementation of gauge invariance. We aim to understand whether it is worth improving the implementation of gauge invariance, by studying whether the current approach can successfully extract the form factor in a toy model. To do this, we must first understand the constraints placed on the amplitude by gauge invariance.

%\begin{itemize}
%\item Define differential cross section
%\item Define kinematic variables
%\end{itemize}

\section{Gauge Invariance and the Ward Green Takahashi Identities}

We are interested in extracting the pion electromagnetic form factor in the intermediate $Q^2$ region. In lieu of an exact solution from QCD, we can attempt to build a model for the interaction. In this case, respecting the symmetries of the fundamental theory is essential. The electromagnetic gauge symmetry is one such symmetry which is exactly respected in pion electro-production. We can thus use this condition to constrain the form of the pion electro-production amplitude. We begin by recalling some basic facts about electromagnetic gauge invariance.

In QED, it is well known that the gauge invariance of the interaction is expressed through relationships between the $n$ and $n+1$ point Green's functions. These identities are collectively referred to as the \textit{Ward Green Takahashi Identities}~\cite{Ward:1950xp,Green:1954zza,Takahashi:1957xn}. In Ref.~\cite{Nishijima1961} Nishijima showed that the Ward Green Takahashi identities are satisfied for a general gauge invariant Lagrangian, independent of the explicit form. The implication of this is that in an effective field theoretic description of pion electro-production, the Ward Green Takahashi Identities from QED must be satisfied. The generic form for these identities in momentum space is
\begin{equation}
q_{\mu} \Gamma_n^{\mu\ldots}(p, \ldots;q) = \Phi_n(\Gamma_{n-1},\Gamma_{n-2},\ldots, \Gamma_2),
\end{equation}
where $q_{\mu}$ is an external boson momentum contracted with the appropriate
Lorentz index of a Green's function. These identities equate this
contraction to a combination of lower Green's functions, denoted symbolically
here by $\Phi_n$. In particular, we are interested in
\begin{equation}
-iq_\mu \Gamma^\mu(p,p^\p)=S_F^{-1}(p^\p)-S_F^{-1}(p),
\end{equation}
and
\begin{equation}
\begin{split}
-iq_\mu \Delta^\mu(q;p_\pi,p^\p,p)=&\Gamma(p_\pi-q,p^\p,p)+\Gamma(p_\pi,p^\p-q,p)  
\\
&-\Gamma(p_\pi,p^\p,p+q),
\end{split}
\end{equation}
where $\Gamma^\mu$ and $S_F(p)$ are the renormalized vertices and propagators, respectively, and $\Delta^\mu$ and $\Gamma$ are the vector four-point and scalar three-point vertices, respectively. 
For a bosonic particle, the most general forms are:
\begin{align}
S_F(p)=&\frac{i}{p^2-m^2-\Sigma(p^2)},
\\
\Gamma^\mu(p,p^\p)=&(p+p^\p)^\mu f_1(p^2,p^{\p2},q^2)+(p-p^\p)^\mu f_2(p^2,p^{\p2},q^2)\,.
\end{align}
The equations for a fermionic particle are more complicated (the most general form of the self energy contains two Lorentz invariant functions, and the most general vertex may be decomposed into twelve Lorentz invariant functions). In this paper, we consider a model in which all particles involved in the interaction are {\it{bosonic}}, so we make no further mention of the fermionic case.

%A partic

Importantly, the Ward Green Takahashi Identities are also satisfied order-by-order in perturbation theory. In this case, the full propagators and vertices are replaced with their approximations, valid at the specific order of perturbation theory being calculated. Not only are they an important check of the model's validity, but gauge invariance is essential in ensuring renormalisability and unitarity of the theory.

%\begin{itemize}
%\item Write a concluding statement, which summarizes the importance of gauge invariance.
%\end{itemize}

%\begin{itemize}
%\item Also need the 4-pt identity
%\item Special case gives overall current conservation $q_\mu \Delta^\mu=0$.
%\end{itemize}

\section{Gauge Invariance in the VGL Model}

%\begin{itemize}
%\item Describe the implementation of gauge invariance in VGL Model.
%\item Point out unnatural factorization.
%\end{itemize}

The Ward Green Takahashi Identities are valid for arbitrary matrix elements. In the limit that the external particles are on their respective mass shells, one can show that the Ward Green Takahashi Identity for the pion electro-production amplitude  reduces to
\begin{equation}
q_\mu\mathcal{M}^\mu=0 \quad.
\end{equation}
In any gauge invariant model, this property must be upheld, to ensure that current conservation has been preserved. One can show that the VGL Model does indeed satisfy this requirement. Since it is central to our discussion of the appropriateness of the pion form factor in the VGL Model amplitude, we will show how gauge invariance is satisfied in this model. To begin with however, we consider the simpler case of the Born Term Model.

\subsection{Gauge Invariance in the Born Term Model}

The Born Term Model is defined by the matrix element 
\begin{equation}
\begin{split}
i\mathcal{M}_{\text{BTM}}^\mu=&\frac{g_A}{\sqrt{2}f_\pi}\overline{u}_N(p^\p,s^\p)\gamma_5\slashed{p}_\pi S_F^N(p_s)(-ie\gamma^\mu)u_N(p,s)
\\
&+\frac{g_A}{\sqrt{2}f_\pi}\overline{u}_N(p^\p,s^\p)\gamma_5\slashed{p}_tu_N(p,s)S_F^\pi(p_t)
\\
&\times (-ie)(p_t+p_\pi)^\mu
\\
&-\frac{g_Ae}{\sqrt{2}f_\pi}\overline{u}_N(p^\p,s^\p)\gamma_5\gamma^\mu u_N(p,s).
\end{split}
\tag{\ref{eq:BTM}}
\end{equation}
We consider contracting $q_\mu$ into this matrix element:
\begin{equation}
\begin{split}
i q_\mu\mathcal{M}_{\text{BTM}}^\mu=&\frac{g_Ae}{\sqrt{2}f_\pi}\overline{u}_N(p^\p,s^\p)\bigg[\gamma_5\slashed{p}_\pi \frac{(\slashed{p}_s+m_N)}{s-m_N^2}\slashed{q}
\\
&+\gamma_5\slashed{p}_t\frac{q\cdot(p_t+p_\pi)}{t-m_\pi^2}
-\gamma_5\slashed{q}\bigg]u_N(p,s).
\end{split}
\end{equation}
After some algebra, one arrives at
\begin{equation}
\begin{split}
i q_\mu\mathcal{M}_\text{BTM}^\mu=&\frac{g_Ae}{\sqrt{2}f_\pi}\overline{u}_N(p^\p,s^\p)\gamma_5\bigg[
\slashed{p}_\pi -\slashed{p}_t-\slashed{q}\bigg]u_N(p,s).
\end{split}
\end{equation}
%-\frac{(q^2-2p^\p.q)}{q^2-2p^\p.q}\slashed{p}_\pi
%
By noting that $p_t=p_\pi-q$, it is possible to see that the hadronic current is conserved. Note that this is true even for $q^2\neq0$, as it must be. Importantly though, tracing the origins of the terms back, it is possible to see that a cancellation occurs between the $t$, $s$ and Kroll-Ruderman terms. In other words, if one wishes to modify the above form of the  Born Term Model, one must do it in such a way that the cancellation persists.

It is reasonably straightforward to show that by multiplying each of the diagrams by a different momentum dependent function, the only possible way to ensure gauge invariance is to set all of these functions equal. In other words, if gauge invariance is preserved for the amplitude $i\mathcal{M}^\mu$, it will also be preserved for the amplitude $f\times[i\mathcal{M}^\mu]$, where $f$ is a general momentum dependent function. It is this result which is essential to understand the VGL Model's implementation of gauge invariance, and is the reason we described the Reggeization of the amplitude a multiplicative operaton. Recalling that it is possible to write the Reggeized amplitude as
\begin{equation}
i\mathcal{M}_{\text{R}}^\mu=S_F^{\pi-1}(p_t)S_R^\pi(p_t)\big[i\mathcal{M}_{\text{BTM}}^\mu\big]
\tag{\ref{eq:regge}},
\end{equation}
it should be clear that the Reggeized amplitude is still gauge invariant. Finally, the structure of the pion is incorporated by multiplying the Reggeized amplitude by the pion form factor $F_\pi(Q^2)$. We thus arrive at the VGL Model matrix element:
\begin{equation}
i\mathcal{M}_\text{VGL}^\mu=i\mathcal{M}_{\text{R}}^\mu F_\pi(Q^2).
\label{eq:vgl}
\end{equation}
By writing the VGL Model amplitude in this form, it is easy to see why gauge invariance is preserved; it is a consequence of the underlying Born Term diagrams which arise from a gauge invariant Lagrangian. This completes our discussion of the VGL Model. In order to determine whether this somewhat unnatural approximation leads to any inconsistencies in the extracted form factor, we will repeat the $F_\pi$ analysis in a simple model whose form factor we can calculate exactly. This will allow us to determine how well one can reconstruct the pion form factor.

\section{A Toy Model of Pion Electro-production}

We have now seen how the VGL Model preserves gauge invariance. In order to determine the consequences for the extracted pion form factor, we will examine how well the approach works in a simple toy model, where we can calculate the form factor and cross section exactly. Our criteria for a suitable model is twofold;

\begin{enumerate}
\item it must be gauge invariant and 
\item the nucleon and pion must have different form factors.
\end{enumerate}

A suitable model for this is Miller's simple model of the nucleon's electromagnetic form factors, described in Ref.~\cite{Miller:2009sg}. In this simple model, we consider a quantum field theory describing the interaction of a scalar `nucleon' and scalar `pion'. To be clear, we define a scalar `nucleon' doublet $\Psi_N$
\begin{equation}
\Psi_N=
\begin{bmatrix}
\psi_p \\
\psi_n 
\end{bmatrix},
\end{equation}
and a scalar `pion' triplet $\boldsymbol{\pi}$
\begin{equation}
\boldsymbol{\pi}=
\begin{bmatrix}
\pi^+ \\
\pi^- \\
\pi^0 
\end{bmatrix},
\end{equation}
where $\pi^+=(\pi_1-i\pi_2)/\sqrt{2}$ and $\pi^-=(\pi_1+i\pi_2)/\sqrt{2}$ we may write the Lagrangian as
\begin{equation}
\begin{split}
\mathcal{L}=&\frac{1}{2}(\partial_\mu\Phi_N)^2-\frac{1}{2}m_N^2\Psi_N^2+\frac{1}{2}(\partial_\mu\boldsymbol{\pi})^2-\frac{1}{2}m_N^2\boldsymbol{\pi}^2
\\
&-g_{\pi N}\Psi_N^\dagger\boldsymbol{\tau}\cdot\boldsymbol{\pi}\Psi_N ,
\end{split}
\end{equation}
where $\boldsymbol{\tau}$ is the isospin vector. Gauging the Lagrangian leads to electromagnetic interactions between the charged particles in the theory. We are interested in (scalar)\\
 $\gamma^*+p\to\pi^++n$. In order to preserve gauge invariance, we calculate one-loop corrections to the tree level cross section. Since gauge invariance is preserved order-by-order in perturbation theory, the resulting theory will certainly be gauge invariant. At one loop, we have 13 diagrams we must evaluate, plus 6 counter terms. We show these diagrams below in Fig.~\ref{fig:5}.

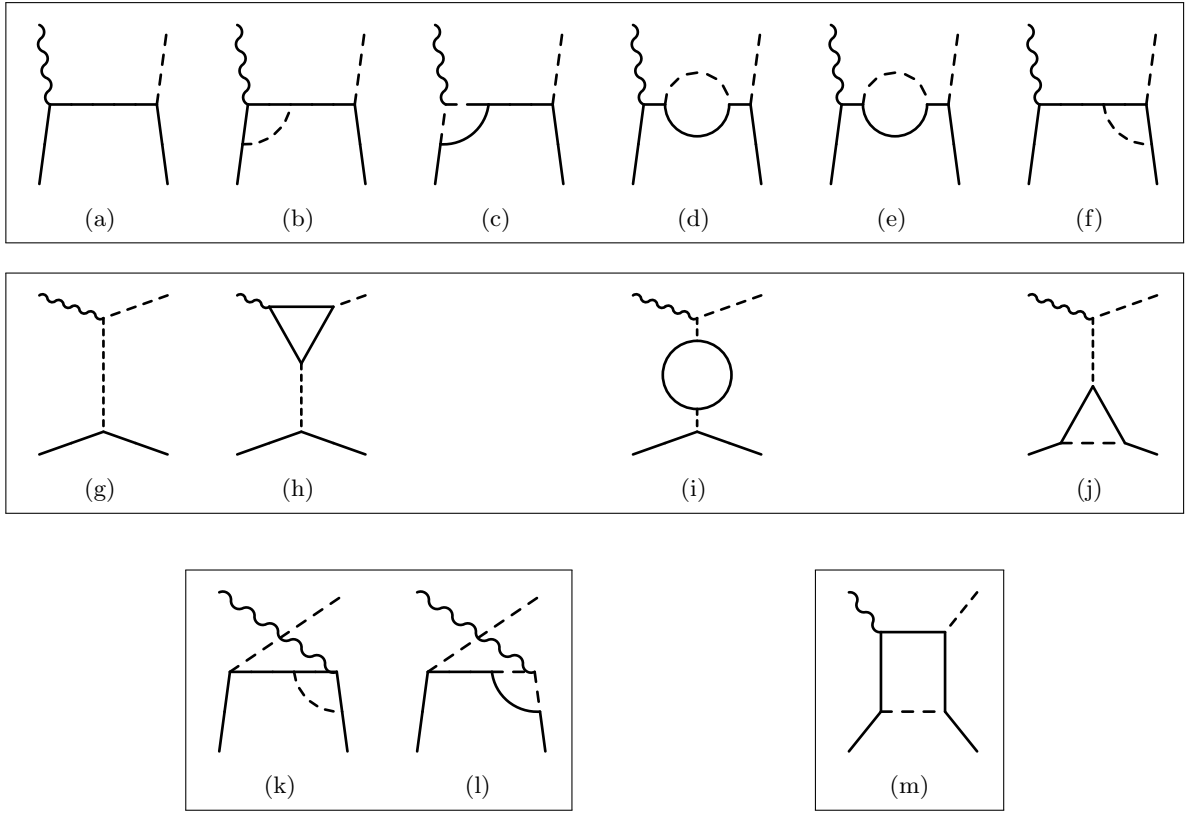
\begin{figure*}[h]
\centering
\fbox{
%% Tree level diagram
\begin{subfigure}{0.11\textwidth}
\vspace{5pt}
\begin{fmfgraph}(60,60)
\fmfleft{l1,l2}
\fmfright{r1,r2}
\fmf{plain,tension=2}{l1,v1,v2}
\fmf{plain}{v2,v3}
\fmf{plain}{v3,v4,v5,v6}
\fmf{plain}{v6,v7}
\fmf{plain,tension=2}{v7,v8,r1}
\fmf{photon}{l2,v2}
\fmf{dashes}{v7,r2}
\fmffreeze
%\fmf{dashes,right=0.4}{v1,v4}
%\fmf{dashes,left=1}{v3,v6}
%\fmf{plain,right=1}{v3,v6}
%\fmf{dashes,right=0.4}{v5,v8}
%\fmfdot{v1,v2,v3,v4,v5,v6,v7,v8,l1,l2,r1,r2}
\end{fmfgraph}
\vspace{-10pt}
\subcaption{}
\end{subfigure}
\hspace{10pt}
\vspace{10pt}
%
% First em vertex correction
\begin{subfigure}{0.11\textwidth}
\vspace{5pt}
\begin{fmfgraph}(60,60)
\fmfleft{l1,l2}
\fmfright{r1,r2}
\fmf{plain,tension=2}{l1,v1,v2}
\fmf{plain}{v2,v3}
\fmf{plain}{v3,v4,v5,v6}
\fmf{plain}{v6,v7}
\fmf{plain,tension=2}{v7,v8,r1}
\fmf{photon}{l2,v2}
\fmf{dashes}{v7,r2}
\fmffreeze
\fmf{dashes,right=0.4}{v1,v4}
%\fmf{dashes,left=1}{v3,v6}
%\fmf{plain,right=1}{v3,v6}
%\fmf{dashes,right=0.4}{v5,v8}
%\fmfdot{v1,v2,v3,v4,v5,v6,v7,v8,l1,l2,r1,r2}
\end{fmfgraph}
\vspace{-10pt}
\subcaption{}
\end{subfigure}
\hspace{10pt}
%
% Second em vertex correction
\begin{subfigure}{0.11\textwidth}
\vspace{5pt}
\begin{fmfgraph}(60,60)
\fmfleft{l1,l2}
\fmfright{r1,r2}
\fmf{plain,tension=2}{l1,v1}
\fmf{dashes,tension=2}{v1,v2}
\fmf{dashes}{v2,v3}
\fmf{plain}{v3,v4,v5,v6}
\fmf{plain}{v6,v7}
\fmf{plain,tension=2}{v7,v8,r1}
\fmf{photon}{l2,v2}
\fmf{dashes}{v7,r2}
\fmffreeze
\fmf{plain,right=0.4}{v1,v4}
%\fmf{dashes,left=1}{v3,v6}
%\fmf{plain,right=1}{v3,v6}
%\fmf{dashes,right=0.4}{v5,v8}
%\fmfdot{v1,v2,v3,v4,v5,v6,v7,v8,l1,l2,r1,r2}
\end{fmfgraph}
\vspace{-10pt}
\subcaption{}
\end{subfigure}
\hspace{10pt}
%
%% First self energy graph (charged pion)
\begin{subfigure}{0.11\textwidth}
\vspace{5pt}
\begin{fmfgraph}(60,60)
\fmfleft{l1,l2}
\fmfright{r1,r2}
\fmf{plain,tension=2}{l1,v1}
\fmf{plain,tension=2}{v1,v2}
\fmf{plain}{v2,v3}
\fmf{phantom}{v3,v4,v5,v6}
\fmf{plain}{v6,v7}
\fmf{plain,tension=2}{v7,v8,r1}
\fmf{photon}{l2,v2}
\fmf{dashes}{v7,r2}
\fmffreeze
%\fmf{plain,right=0.4}{v1,v4}
\fmf{dashes,left=1}{v3,v6}
\fmf{plain,right=1}{v3,v6}
%\fmf{plain,right=1}{v3,v6}
%\fmf{dashes,right=0.4}{v5,v8}
%\fmfdot{v1,v2,v3,v4,v5,v6,v7,v8,l1,l2,r1,r2}
\end{fmfgraph}
\vspace{-10pt}
\subcaption{}
\end{subfigure}
\hspace{10pt}
%
%% Second self energy graph (neutral pion)
\begin{subfigure}{0.11\textwidth}
\vspace{5pt}
\begin{fmfgraph}(60,60)
\fmfleft{l1,l2}
\fmfright{r1,r2}
\fmf{plain,tension=2}{l1,v1}
\fmf{plain,tension=2}{v1,v2}
\fmf{plain}{v2,v3}
\fmf{phantom}{v3,v4,v5,v6}
\fmf{plain}{v6,v7}
\fmf{plain,tension=2}{v7,v8,r1}
\fmf{photon}{l2,v2}
\fmf{dashes}{v7,r2}
\fmffreeze
%\fmf{plain,right=0.4}{v1,v4}
\fmf{dashes,left=1}{v3,v6}
\fmf{plain,right=1}{v3,v6}
%\fmf{plain,right=1}{v3,v6}
%\fmf{dashes,right=0.4}{v5,v8}
%\fmfdot{v1,v2,v3,v4,v5,v6,v7,v8,l1,l2,r1,r2}
\end{fmfgraph}
\vspace{-10pt}
\subcaption{}
\end{subfigure}
\hspace{10pt}
%
% Strong Vertex Correction
\begin{subfigure}{0.11\textwidth}
\vspace{5pt}
\begin{fmfgraph}(60,60)
\fmfleft{l1,l2}
\fmfright{r1,r2}
\fmf{plain,tension=2}{l1,v1}
\fmf{plain,tension=2}{v1,v2}
\fmf{plain}{v2,v3}
\fmf{plain}{v3,v4,v5,v6}
\fmf{plain}{v6,v7}
\fmf{plain,tension=2}{v7,v8,r1}
\fmf{photon}{l2,v2}
\fmf{dashes}{v7,r2}
\fmffreeze
%\fmf{plain,right=0.4}{v1,v4}
%\fmf{dashes,left=1}{v3,v6}
%\fmf{plain,right=1}{v3,v6}
\fmf{dashes,right=0.4}{v5,v8}
%\fmfdot{v1,v2,v3,v4,v5,v6,v7,v8,l1,l2,r1,r2}
\end{fmfgraph}
\vspace{-10pt}
\subcaption{}
\end{subfigure}
}
\hspace{10pt}
\vspace{10pt}

\fbox{
% t-channel tree level diagram
\begin{subfigure}{0.11\textwidth}
\vspace{5pt}
\begin{fmfgraph}(60,60)
\fmfleft{b1,t1}
\fmfright{b2,t2}
\fmf{dashes}{t2,v1,v2}
\fmf{dashes}{v2,v3}
\fmf{dashes}{v3,v4,v5,v6}
\fmf{dashes}{v6,v7}
\fmf{plain}{v7,v8,b2}
\fmf{photon}{t1,v2a,v2}
\fmf{plain}{v7,v7a,b1}
\fmffreeze
%\fmf{dashes}{v1,v4}
%\fmf{dashes,left=1}{v3,v6}
%\fmf{plain,right=1}{v3,v6}
%\fmf{dashes}{v5,v8}
%\fmf{plain}{v7a,v2a,v1,v8,v7a}
%\fmfdot{v1,v2,v3,v4,v5,v6,v7,v8,v2a,v7a,t1,t2,b1,b2}
\end{fmfgraph}
\vspace{-10pt}
\subcaption{}
\end{subfigure}
\hspace{10pt}
%
% vertex correction
\begin{subfigure}{0.11\textwidth}
\vspace{5pt}
\begin{fmfgraph}(60,60)
\fmfleft{b1,t1}
\fmfright{b2,t2}
\fmf{dashes}{t2,v1}
\fmf{phantom}{v1,v2}
\fmf{phantom}{v2,v3,v4}
\fmf{dashes}{v4,v5,v6}
\fmf{dashes}{v6,v7}
\fmf{plain}{v7,v8,b2}
\fmf{photon}{t1,v2a}
\fmf{phantom}{v2a,v2}
\fmf{plain}{v7,v7a,b1}
\fmffreeze
\fmf{plain}{v1,v2a,v4,v1}
%\fmf{dashes}{v1,v4}
%\fmf{dashes,left=1}{v3,v6}
%\fmf{plain,right=1}{v3,v6}
%\fmf{dashes}{v5,v8}
%\fmf{plain}{v7a,v2a,v1,v8,v7a}
%\fmfdot{v1,v2,v3,v4,v5,v6,v7,v8,v2a,v7a,t1,t2,b1,b2}
\end{fmfgraph}
\vspace{-10pt}
\subcaption{}
\end{subfigure}
\hspace{10pt}
%
% BLANK
\begin{subfigure}{0.11\textwidth}
\vspace{5pt}
\begin{fmfgraph}(60,60)
\fmfleft{b1,t1}
\fmfright{b2,t2}
\fmf{phantom}{t2,v1,v2}
\fmf{phantom}{v2,v3}
\fmf{phantom}{v3,v4,v5,v6}
\fmf{phantom}{v6,v7}
\fmf{phantom}{v7,v8,b2}
\fmf{phantom}{t1,v2a,v2}
\fmf{phantom}{v7,v7a,b1}
\fmffreeze
%\fmf{dashes}{v1,v4}
%\fmf{dashes,left=1}{v3,v6}
%\fmf{plain,right=1}{v3,v6}
%\fmf{dashes}{v5,v8}
%\fmf{plain}{v7a,v2a,v1,v8,v7a}
%\fmfdot{v1,v2,v3,v4,v5,v6,v7,v8,v2a,v7a,t1,t2,b1,b2}
\end{fmfgraph}
\vspace{-10pt}
\end{subfigure}
\hspace{10pt}
%
% Self energy
\begin{subfigure}{0.11\textwidth}
\vspace{5pt}
\begin{fmfgraph}(60,60)
\fmfleft{b1,t1}
\fmfright{b2,t2}
\fmf{dashes}{t2,v1,v2}
\fmf{dashes}{v2,v3}
\fmf{phantom}{v3,v4,v5,v6}
\fmf{dashes}{v6,v7}
\fmf{plain}{v7,v8,b2}
\fmf{photon}{t1,v2a,v2}
\fmf{plain}{v7,v7a,b1}
\fmffreeze
%\fmf{dashes}{v1,v4}
\fmf{plain,left=1}{v3,v6}
\fmf{plain,right=1}{v3,v6}
%\fmf{dashes}{v5,v8}
%\fmf{plain}{v7a,v2a,v1,v8,v7a}
%\fmfdot{v1,v2,v3,v4,v5,v6,v7,v8,v2a,v7a,t1,t2,b1,b2}
\end{fmfgraph}
\vspace{-10pt}
\subcaption{}
\end{subfigure}
\hspace{10pt}
%
% BLANK
\begin{subfigure}{0.11\textwidth}
\vspace{5pt}
\begin{fmfgraph}(60,60)
\fmfleft{b1,t1}
\fmfright{b2,t2}
\fmf{phantom}{t2,v1,v2}
\fmf{phantom}{v2,v3}
\fmf{phantom}{v3,v4,v5,v6}
\fmf{phantom}{v6,v7}
\fmf{phantom}{v7,v8,b2}
\fmf{phantom}{t1,v2a,v2}
\fmf{phantom}{v7,v7a,b1}
\fmffreeze
%\fmf{dashes}{v1,v4}
%\fmf{dashes,left=1}{v3,v6}
%\fmf{plain,right=1}{v3,v6}
%\fmf{dashes}{v5,v8}
%\fmf{plain}{v7a,v2a,v1,v8,v7a}
%\fmfdot{v1,v2,v3,v4,v5,v6,v7,v8,v2a,v7a,t1,t2,b1,b2}
\end{fmfgraph}
\vspace{-10pt}
\end{subfigure}
\hspace{10pt}
%
% Strong vertex correction
\begin{subfigure}{0.11\textwidth}
\vspace{5pt}
\begin{fmfgraph}(60,60)
\fmfleft{b1,t1}
\fmfright{b2,t2}
\fmf{dashes}{t2,v1,v2}
\fmf{dashes}{v2,v3}
\fmf{dashes}{v3,v4,v5}
\fmf{phantom}{v5,v6,v7}
\fmf{phantom}{v7,v8}
\fmf{plain}{v8,b2}
\fmf{photon}{t1,v2a,v2}
\fmf{phantom}{v7,v7a}
\fmf{plain}{v7a,b1}
\fmffreeze
\fmf{plain}{v7a,v5,v8}
\fmf{dashes}{v7a,v8}
%\fmf{dashes}{v1,v4}
%\fmf{dashes,left=1}{v3,v6}
%\fmf{plain,right=1}{v3,v6}
%\fmf{dashes}{v5,v8}
%\fmf{plain}{v7a,v2a,v1,v8,v7a}
%\fmfdot{v1,v2,v3,v4,v5,v6,v7,v8,v2a,v7a,t1,t2,b1,b2}
\end{fmfgraph}
\vspace{-10pt}
\subcaption{}
\end{subfigure}
}
\hspace{10pt}
\vspace{20pt}

% BLANK
\begin{subfigure}{0.11\textwidth}
\vspace{5pt}
\begin{fmfgraph}(60,60)
\fmfleft{b1,t1}
\fmfright{b2,t2}
\fmf{phantom}{t2,v1,v2}
\fmf{phantom}{v2,v3}
\fmf{phantom}{v3,v4,v5,v6}
\fmf{phantom}{v6,v7}
\fmf{phantom}{v7,v8,b2}
\fmf{phantom}{t1,v2a,v2}
\fmf{phantom}{v7,v7a,b1}
\fmffreeze
%\fmf{dashes}{v1,v4}
%\fmf{dashes,left=1}{v3,v6}
%\fmf{plain,right=1}{v3,v6}
%\fmf{dashes}{v5,v8}
%\fmf{plain}{v7a,v2a,v1,v8,v7a}
%\fmfdot{v1,v2,v3,v4,v5,v6,v7,v8,v2a,v7a,t1,t2,b1,b2}
\end{fmfgraph}
\vspace{-10pt}
\end{subfigure}
\hspace{10pt}
\fbox{
% Strong Vertex Correction
\begin{subfigure}{0.11\textwidth}
\vspace{5pt}
\begin{fmfgraph}(60,60)
\fmfleft{l1,l2}
\fmfright{r1,r2}
\fmf{plain,tension=2}{l1,v1}
\fmf{plain,tension=2}{v1,v2}
\fmf{plain}{v2,v3}
\fmf{plain}{v3,v4,v5,v6}
\fmf{plain}{v6,v7}
\fmf{plain,tension=2}{v7,v8,r1}
\fmf{phantom}{l2,v2}
\fmf{phantom}{v7,r2}
\fmffreeze
\fmf{photon}{l2,v7}
\fmf{dashes}{v2,r2}
\fmf{dashes,right=0.4}{v5,v8}
\end{fmfgraph}
\vspace{-10pt}
\subcaption{}
\end{subfigure}
\hspace{10pt}
%
% Strong Vertex Correction
\begin{subfigure}{0.11\textwidth}
\vspace{5pt}
\begin{fmfgraph}(60,60)
\fmfleft{l1,l2}
\fmfright{r1,r2}
\fmf{plain,tension=2}{l1,v1}
\fmf{plain,tension=2}{v1,v2}
\fmf{plain}{v2,v3}
\fmf{plain}{v3,v4,v5}
\fmf{dashes}{v5,v6}
\fmf{dashes}{v6,v7}
\fmf{dashes,tension=2}{v7,v8}
\fmf{plain,tension=2}{v8,r1}
\fmf{phantom}{l2,v2}
\fmf{phantom}{v7,r2}
\fmffreeze
\fmf{photon}{l2,v7}
\fmf{dashes}{v2,r2}
\fmf{plain,right=0.4}{v5,v8}
\end{fmfgraph}
\vspace{-10pt}
\subcaption{}
\end{subfigure}
}
\hspace{10pt}
% BLANK
\begin{subfigure}{0.11\textwidth}
\vspace{5pt}
\begin{fmfgraph}(60,60)
\fmfleft{b1,t1}
\fmfright{b2,t2}
\fmf{phantom}{t2,v1,v2}
\fmf{phantom}{v2,v3}
\fmf{phantom}{v3,v4,v5,v6}
\fmf{phantom}{v6,v7}
\fmf{phantom}{v7,v8,b2}
\fmf{phantom}{t1,v2a,v2}
\fmf{phantom}{v7,v7a,b1}
\fmffreeze
%\fmf{dashes}{v1,v4}
%\fmf{dashes,left=1}{v3,v6}
%\fmf{plain,right=1}{v3,v6}
%\fmf{dashes}{v5,v8}
%\fmf{plain}{v7a,v2a,v1,v8,v7a}
%\fmfdot{v1,v2,v3,v4,v5,v6,v7,v8,v2a,v7a,t1,t2,b1,b2}
\end{fmfgraph}
\vspace{-10pt}
\end{subfigure}
\hspace{10pt}
%
% Strong vertex correction
\fbox{
\begin{subfigure}{0.11\textwidth}
\vspace{5pt}
\begin{fmfgraph}(60,60)
\fmfleft{b1,t1}
\fmfright{b2,t2}
\fmf{plain,tension=2}{v1,b1}
\fmf{photon,tension=2}{t1,v2}
\fmf{dashes,tension=2}{t2,v3}
\fmf{plain,tension=2}{v4,b2}
\fmf{plain}{v1,v2,v3,v4}
\fmf{dashes}{v1,v4}
%\fmfdot{v1,v2,v3,v4,v5,v6,v7,v8,v2a,v7a,t1,t2,b1,b2}
\end{fmfgraph}
\vspace{-10pt}
\subcaption{}
\end{subfigure}
}
\hspace{10pt}
%
% BLANK
\begin{subfigure}{0.11\textwidth}
\vspace{5pt}
\begin{fmfgraph}(60,60)
\fmfleft{b1,t1}
\fmfright{b2,t2}
\fmf{phantom}{t2,v1,v2}
\fmf{phantom}{v2,v3}
\fmf{phantom}{v3,v4,v5,v6}
\fmf{phantom}{v6,v7}
\fmf{phantom}{v7,v8,b2}
\fmf{phantom}{t1,v2a,v2}
\fmf{phantom}{v7,v7a,b1}
\fmffreeze
%\fmf{dashes}{v1,v4}
%\fmf{dashes,left=1}{v3,v6}
%\fmf{plain,right=1}{v3,v6}
%\fmf{dashes}{v5,v8}
%\fmf{plain}{v7a,v2a,v1,v8,v7a}
%\fmfdot{v1,v2,v3,v4,v5,v6,v7,v8,v2a,v7a,t1,t2,b1,b2}
\end{fmfgraph}
\vspace{-10pt}
\end{subfigure}
\hspace{10pt}
\vspace{10pt}

\caption{Diagrams contributing to $\gamma^*+p\to\pi+n$. We list first all $s$-channel diagrams, all $t$-channel diagrams, both $u$-channel diagrams and the single box diagram. }\label{fig:5}
\end{figure*}

Explicit expressions for these diagrams are given in \ref{app:diagrams}. Since we are calculating this model at one loop order, divergences appear which we must absorb into the definitions of our couplings and masses. We use the on-shell renormalization scheme:
\begin{align}
\Sigma(p^2)\big|_{p^2=m^2}=&0  \quad,
\\
\frac{d}{dp^2}\Sigma(p^2)\bigg|_{p^2=m^2}=&0  \quad,
\\
\lim_{q\to0}(-ie)\Gamma^\mu(p,p^\p)=&(-ie)2p^\mu   \quad.
\end{align}
Amaldi, Fubini and Furlan in Ref.~\cite{Amaldi:1979vh} provided the details of the relationship between the hadronic matrix element and the differential cross section, decomposed in terms of the longitudinal, transverse and interference terms. We use the Mathematica package \textsc{FeynCalc}~\cite{Mertig:1990an,Shtabovenko:2016sxi} to determine our final expression for these structure functions, and then perform the loop integrals using \textsc{QCDLoop}~\cite{Ellis:2007qk}.

\subsection{Form Factors in the Toy Model}

As a first check, we can examine the form factors generated by the loop corrections in this model. As one might predict, the corrections to the form factors generated by the inclusion of the loop diagrams are quite small. Since we are requiring the one-loop diagrams to contribute \textit{all} the $Q^2$ behavior, this is problematic. In order to rectify this, we have chosen to change the coupling which controls the strength of the loop corrections, as well as the masses of the particles propagating in the loops. In other words, we take $m_N\to m_N^\p$, $m_\pi\to m_\pi^\p$ and $g_{\pi N}\to g_{\pi N}^\p$ in the loop integrals only. 

Clearly from the point of view of a consistent quantum field theoretic calculation, this approach is incorrect. Note however that since we do this consistently to each loop diagram, we preserve gauge invariance in this approach. As we are interested in the toy model only from the point of view of determining how well one may extract the pion form factor, rather than attempting to produce a fully consistent calculation, we believe the results are qualitatively meaningful.

\begin{figure}[H]
\centering
\includegraphics[scale=0.4]{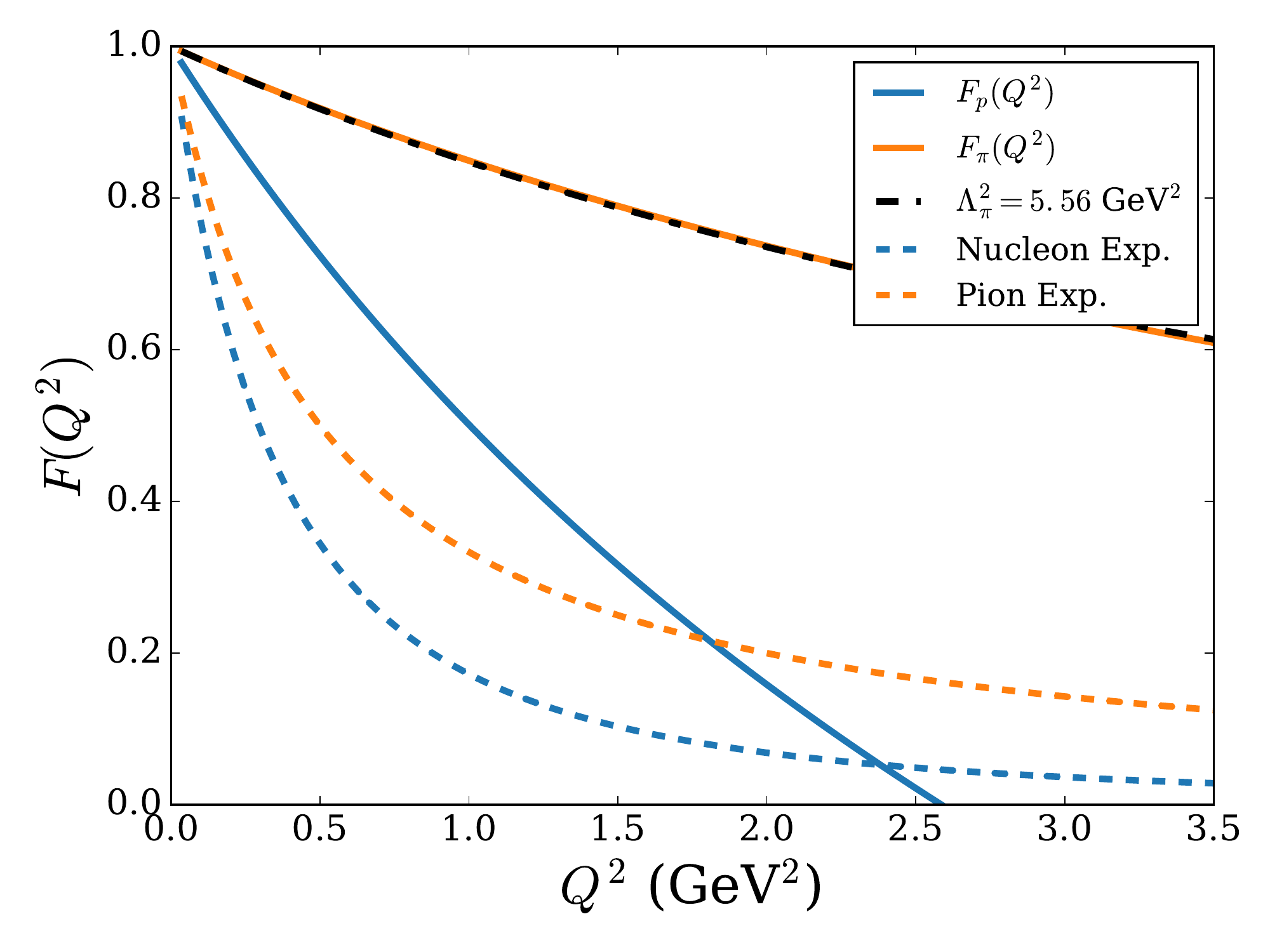}
\caption{Comparison of predicted electromagnetic form factors with parameterizations of the pion and nucleon form factors from data. In principal, there is also a neutron form factor, but due to the chosen mass parameters ($m_N^\p\approx m_\pi^\p$), the neutron form factor is approximately zero (see \ref{app:diagrams} for details). We show the fitted monopole form factor. The agreement between the true pion form factor and the monopole form factor is excellent.}\label{fig:eFF}
\end{figure}

Our chosen parameters are given in Table~\ref{tab:massAndCouplings}. We select parameters to ensure a reasonable separation between $F_p$ and $F_\pi$ and also so that $F_\pi$ falls off slower than $F_p$, as occurs in nature. Since we wish to describe the pion form factor with a monopole form factor, it is important to check that this is a good approximation. We find that the model form factor is well described for a monopole mass parameter of $\Lambda_\pi^2=5.56$ GeV$^2$. This fit is shown in Fig.~\ref{fig:eFF}.

With the free parameters in our model chosen, we may proceed to calculate the cross section and attempt to extract the model pion form factor.

%\begin{itemize}
%\item Need to argue why our results are meaningful.
%\end{itemize}

\begin{table}[H]
\centering
\begin{tabular}{ c  c  c  c  c  c }
\hline
$g_{\pi N}$ & $m_N$ & $m_\pi$ & $g_{\pi N}^\p$ & $m_N^\p$ & $m_\pi^\p$  \\ \hline
1.4 & 0.94 & 0.14 & 20 & 0.7 & 0.71 \\ \hline
\end{tabular}
\caption{Tree level and loop parameters used in this study. All parameters are in units of GeV.}\label{tab:massAndCouplings}
\end{table}

%\begin{figure}[h]
%\includegraphics[scale=0.4]{graphics/plots/agreementFF}
%\caption{Comparison of pion form factor }\label{fig:agreementFF}
%\end{figure}

\section{Extraction of Pion Form Factor}

\begin{figure}[h]
\centering
\includegraphics[scale=0.40]{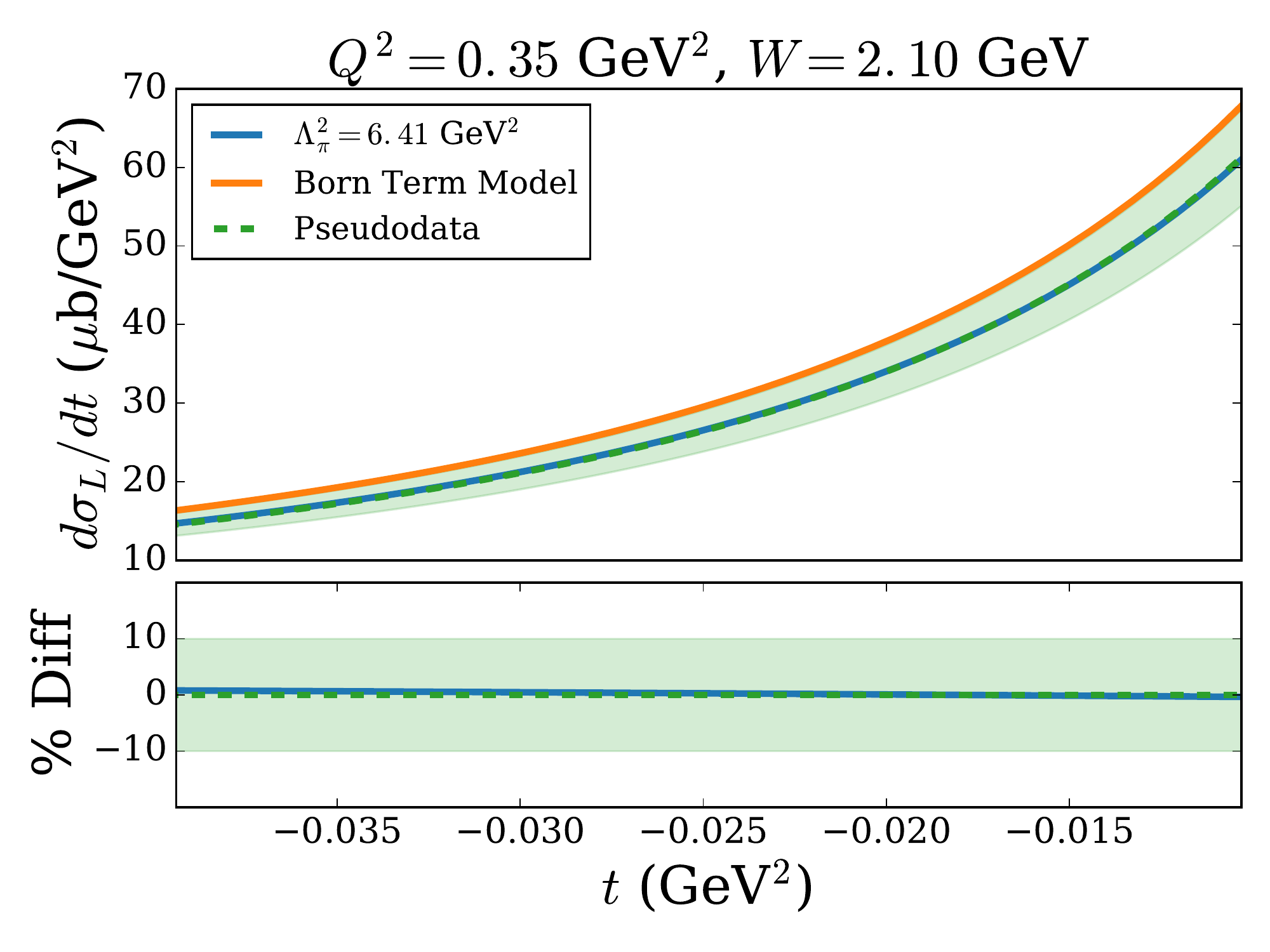}
\caption{(Colour online) Fitting simplified model of cross section to model cross section. Both the pseudodata (dashed green) and simplified model (blue) sit over one another. The extracted $\Lambda_\pi^2$ is related to the extracted pion from factor via $F_\pi(Q^2)=(1+Q^2/\Lambda_\pi^2)^{-1}$.}\label{fig:crossSectionExample}
\end{figure}

\begin{figure}[h]
\centering
\includegraphics[scale=0.40]{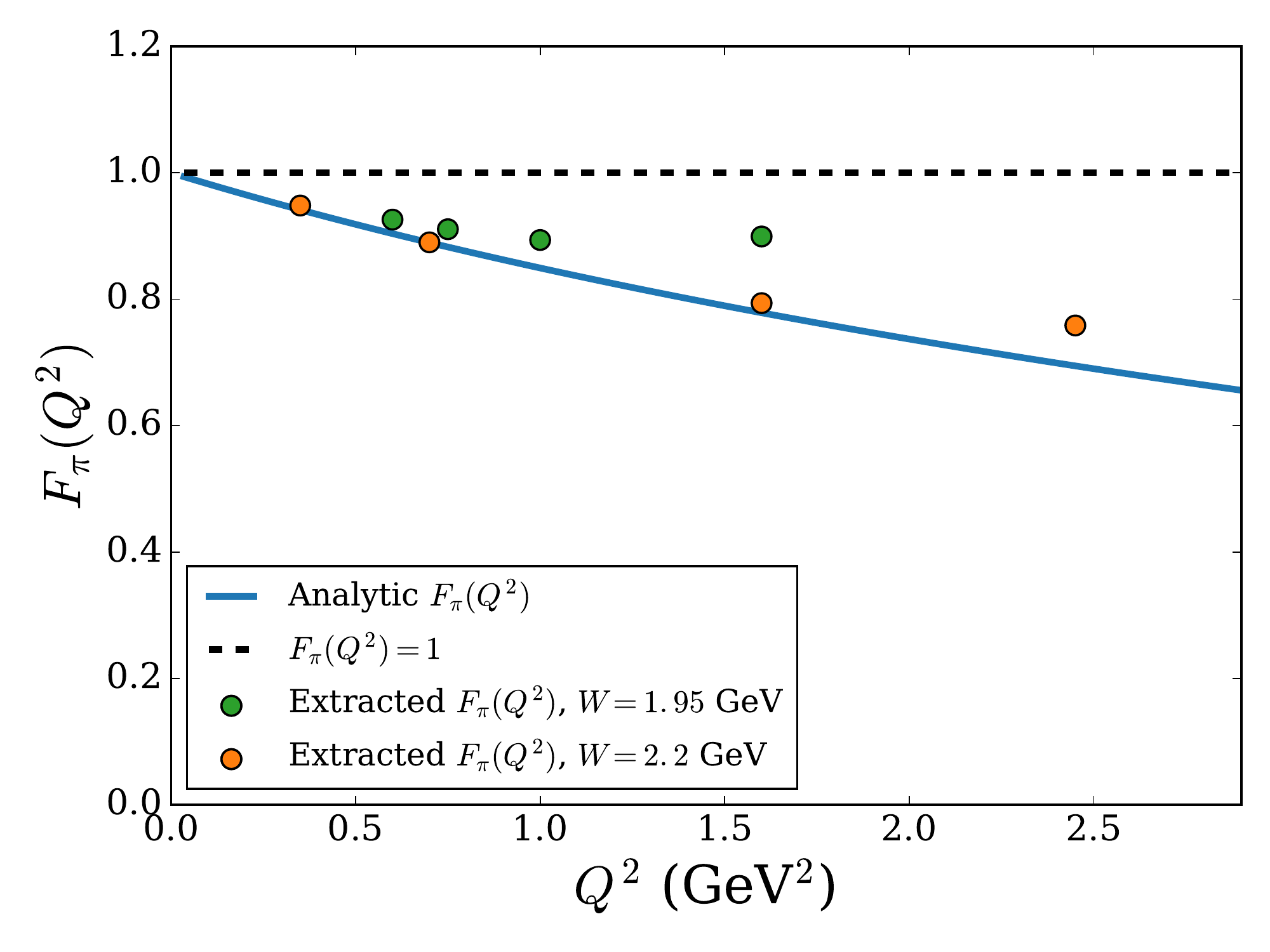}
\caption{(Colour online) Extracted $F_\pi$ in our Toy Model, compared with true model form factor. }\label{fig:extractedFormFactor}
\end{figure}

\begin{figure}[h]
\centering
\includegraphics[scale=0.40]{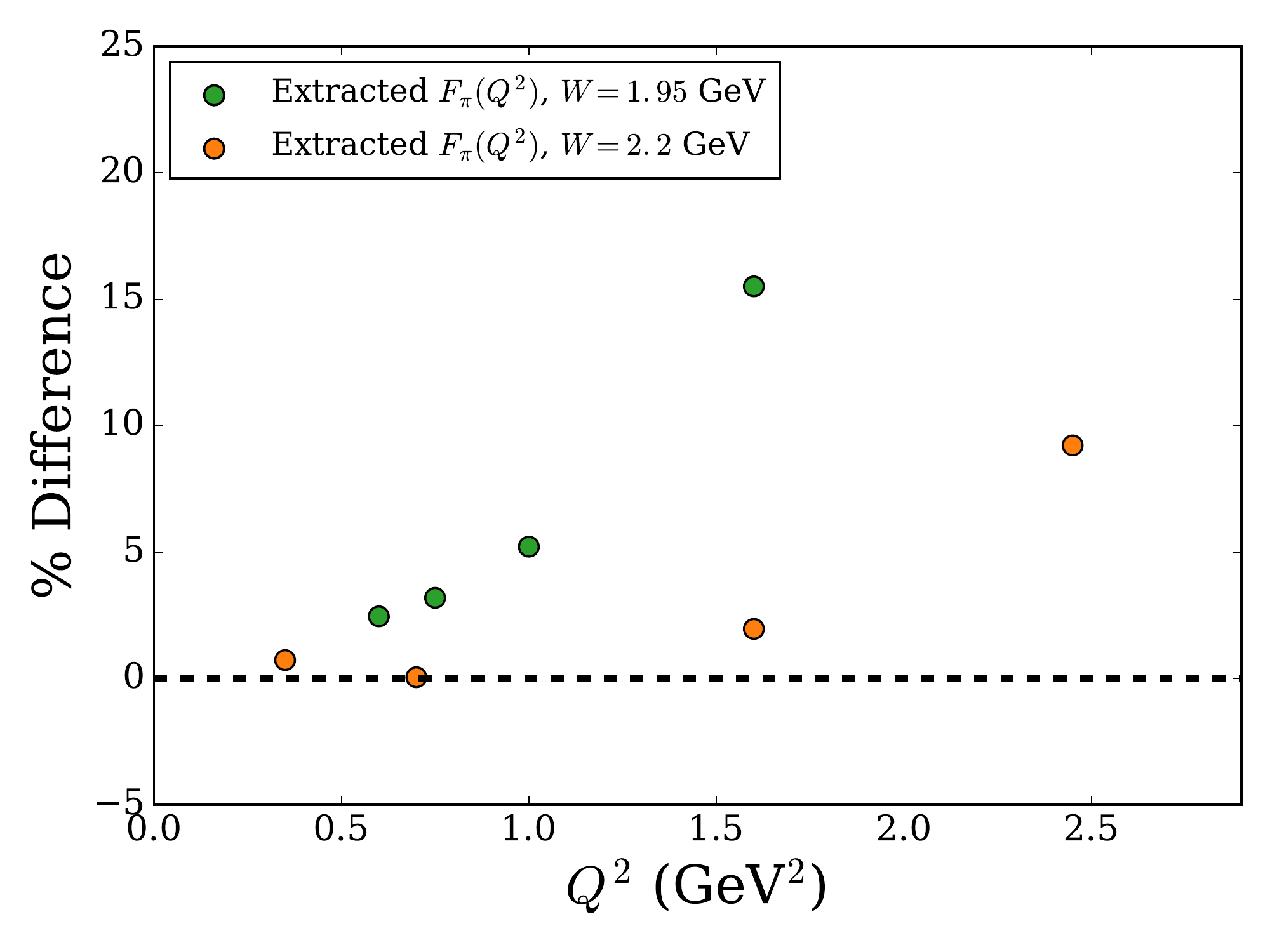}
\caption{(Colour online) Percentage difference between our extracted form factor and the true model form factor. Note that a positive difference corresponds to an overestimation of $F_\pi$. Thus for the kinematic points surveyed, the extracted form factor is overestimated.}
\label{fig:percent_diff}
\end{figure}

The $F_\pi$ Collaboration reports the pion electromagnetic form factor for eight kinematic points, so in our first analysis, we attempt to extract those same points (see Table~\ref{table:kin_points}). We follow a simplified version of their analysis outlined above in Sec.~\ref{sec:fittingVGL}. We outline the steps of the analysis here:

\begin{enumerate}
\item We calculate the loop corrected cross section, with the form factors described in previous section. This cross section is called \textit{pseudodata} in the following step. The model pion form factor, $F_\pi(Q^2)$ shown in Fig.~\ref{fig:eFF} is extracted from this cross section.
\item We generate pseudodata for a range of $t$ values for fixed $Q^2$ and $W$ (dashed green line in Fig.~\ref{fig:crossSectionExample}). As with the $F_\pi$ Collaboration, we choose the range of $t$ to start near the minimum allowed value for the chosen kinematics. Specifically, the cross section is calculated between the minimum and maximum values of $t$ measured (see Ref.~\cite{Blok:2008jy} for explicit values). 
\item We define our model to be the tree level matrix element, and incorporate the pion form factor as a multiplicative factor to the amplitude. This mirrors the approach in the VGL Model. Thus our matrix element is
\begin{equation}
i\mathcal{M}^\mu=i\mathcal{M}_\text{BTM}^\mu F_\pi(Q^2).
\end{equation}
\item We fit our model to the pseudodata to obtain our best fit for the parameter $\Lambda_\pi^2$. This value of $\Lambda_\pi^2$ corresponds to the extracted pion form factor (solid blue line in Fig.~\ref{fig:crossSectionExample}).
\item We plot the resulting extracted pion form factors (see Figs.~\ref{fig:extractedFormFactor},~\ref{fig:percent_diff}).
\end{enumerate}

%\multicolumn{2}{c}{$-t$ range (GeV$^2$)}

\begin{table}[h]
\centering
\begin{tabular}{ c  c  c  c }
\hline
$Q^2$ (GeV$^2$) & $W$ (GeV) & $|t_\text{low}|$ (GeV$^2$) & $|t_\text{high}|$ (GeV$^2$) \\ \hline
0.35 & 2.10 & 0.010 & 0.040 \\
0.60 & 1.95 & 0.025 & 0.074 \\
0.70 & 2.19 & 0.030 & 0.250 \\
0.75 & 1.95 & 0.037 & 0.093 \\
1.00 & 1.95 & 0.060 & 0.140  \\
1.60 & 1.95 & 0.135 & 0.255 \\
1.60 & 2.22 & 0.079 & 0.215 \\
2.45 & 2.22 & 0.145 & 0.365 \\ \hline
\end{tabular}
\caption{The $F_\pi$ Collaboration extracted the pion electromagnetic form factor at 8 kinematic points. We will attempt to extract the same 8 kinematic points.}
\label{table:kin_points}
\end{table}

\section{Discussion of the Results}

Examining the fitted model cross sections shown in \\
\ref{app:crossSections}, we can see that the agreement of the fitted model cross section when compared with pseudodata decreases slightly as we go to larger $Q^2$. We note however, that with the exception of the $(Q^2,W)=(1.6,1.95)$ and $(2.45,2.22)$ kinematics, the disagreement between the model and pseudodata is less than ten percent (see Fig.~\ref{fig:percent_diff}). Given the current experimental uncertainties are of this order, we conclude that the VGL model implementaion of gauge invariance should model the cross section reasonably well over the kinematic range examined. This conclusion is borne out by the experimental data in Ref~\cite{Huber:2008id}.

\begin{figure}
\centering
\includegraphics[scale=0.40]{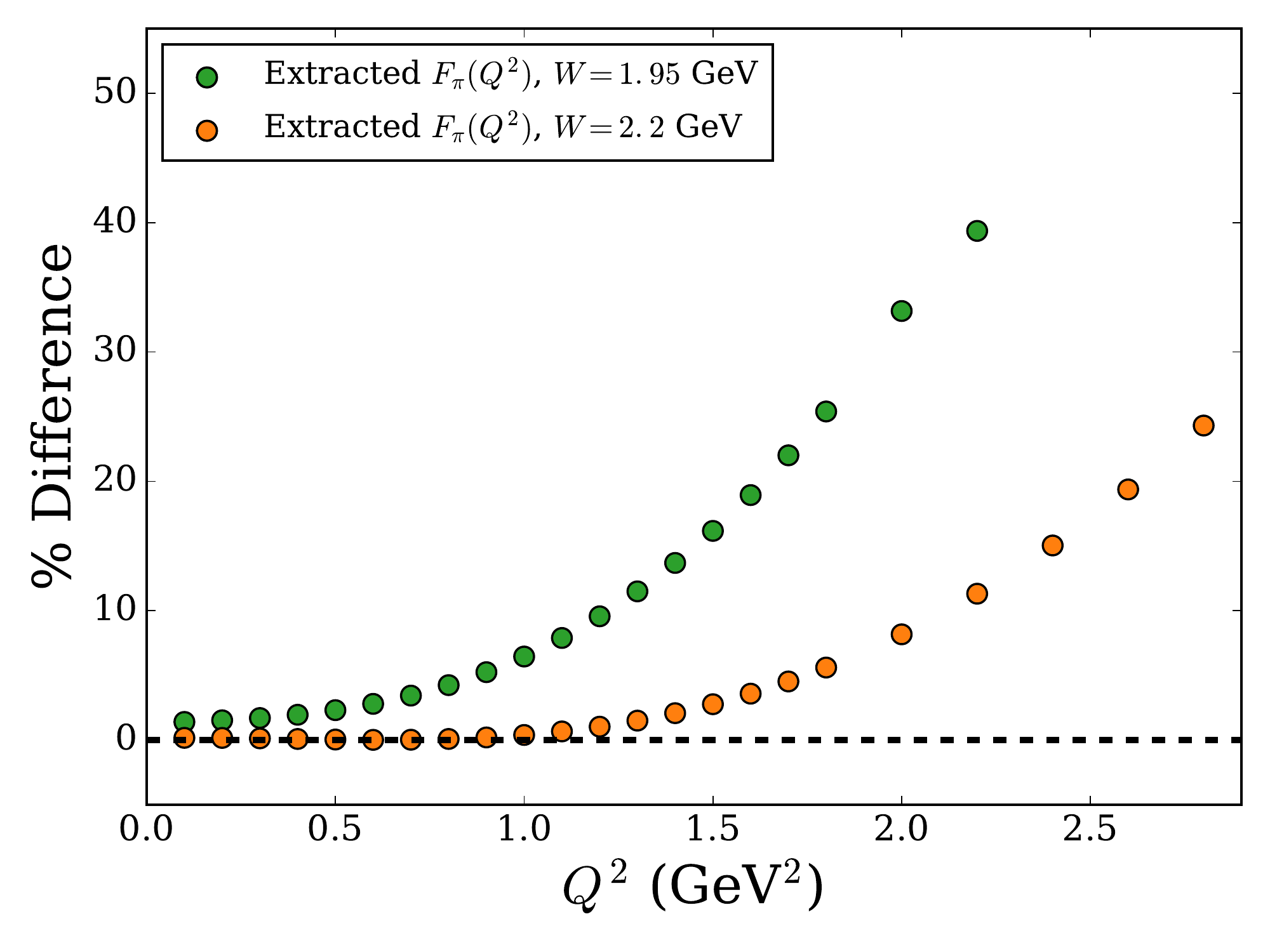}
\caption{(Colour online) Extracting the pion form factor near the minimum kinematically allowed $t$ value. Note that the agreement between the true form factor and the extracted form factor worsens sooner for the $W=1.95$  GeV data, because $|t_\text{min}|$ is larger. Thus the pion photon interaction looks less like the on-shell pion electromagnetic form factor.}\label{fig:percentDiffmint}
\end{figure}

At low momentum transfer, we find that our extracted form factor is in good agreement with the true form factor in the toy model, although in general, we find a better agreement for data points extracted at larger $W$. As the momentum transfer increases, our extracted form factor tends to become a slightly worse representation of the true form factor. In particular, we note from Fig.~\ref{fig:percentDiffmint}, that the extracted form factor appears to trend away from the true form factor.

As noted in Ref.~\cite{Blok:2008jy}, the smallest kinematically allowed absolute value of $t$, denoted $|t_\text{min}|$ may be reduced by measuring at larger $W$, or at smaller $Q^2$. This is important, as this reduces the distance that one has to extrapolate to in order to reach the pion pole. In other words, for smaller absolute value of $t$, the pion photon interaction which occurs in the $t$-channel looks more like the pion electromagnetic form factor measured in elastic $e^-+\pi^+$ scattering. 

To verify our explanation, we extracted the model form factor at $W=1.95$ GeV and $W=2.2$ GeV, for a range of $Q^2$ between 0 and 3 GeV$^2$, using the method outline above. The experimental data approximately spans the first five percent of the allowed $t$ range. We therefore attempt to fit our model cross section to the pseudodata over the first five percent of the allowed $t$ kinematic range for the chosen $Q^2$ and $W$.

%\begin{itemize}
%\item Check what the condition is for fitting a region of the form factor.
%\end{itemize}

 The results of this process are shown in Fig~\ref{fig:percentDiffmint}. As predicted, the agreement between the extracted form factor and the model form factor are good for the large range of $Q^2$ when the $W=2.2$ GeV data is used. This data clearly shows the way the model form factor is being systematically overestimated for increasing $Q^2$. 

It is interesting to speculate about the way this result could carry over to the extraction of the real pion form factor from real data. Indeed, if the relation between the extracted and true pion form factor remained quantitatively the same, this would suggest (experimental uncertainties notwithstanding) that the extracted pion form factor values are currently overestimated. This effect - if observed - would imply that the `true' pion form factor was smaller, bringing the extracted pion form factor closer to the asymptotic limit predicted by perturbative QCD.

%We have coloured points 

%\section{Examining the $W$ Dependence of Model}

%\begin{itemize}
%\item Inspired by Miller.
%\item Want to examine whether the current method can truly extract form factor.
%\item Wanted a model which:
%\begin{itemize}
%\item Preserved gauge invariance
%\item Had different form factors for pion and nucleon.
%\end{itemize}
%\item Lagrangian
%\item Feynman Diagrams
%\item Form factors produced
%\item Summary of extraction method used in J-Lab Paper
%\item Plot of J-Lab Form Factor?
%\item Plot of selected extracted cross section
%\item Plot of extracted form factor
%\end{itemize}

\section{Conclusion}

We began by discussing the theoretical drawbacks with the implementation of gauge invariance in the VGL Model. In particular, we discussed the unnatural factorization of the pion form factor from the matrix element. We proposed a simple toy model which we used to generate pseudodata for pion electro-production. We followed the $F_\pi$ Collaboration's approach to extract the toy model's pion form factor, although we did not extrapolate to the minimum $t$ value. The extracted pion form factor was compared to the true form factor, and it was  found that our extracted form factor was in all cases \textit{larger} than the true form factor. If this result were to hold in the extraction of the experimental pion form factor, it would suggest that the extracted pion form factor is currently overestimated.

Although the present work was simplified by using only bosonic fields, we believe that the results may provide guidance concerning the role of a full implementation of gauge invariance in the physical system. The extension of the present work to the case of fermions is clearly a high priority.

\section*{Acknowledgments}

We thank R. Young, G. Miller and J. Zanotti for helpful discussions. This work was supported by the University of Adelaide and by the Australian Research Council through the ARC Centre of Excellence for Particle Physics at the Terascale (CE110001104) and Discovery Projects DP150103101 and DP180100497.

\appendix
\section{Scalar and Vector Loop Integrals}

We define the following loop integrals. $q_i$ are the external particle momenta, all understood to be entering the diagram. Thus conservation of momentum is $q_1+q_2+q_3+q_4=0$. In the case of the vector three- and four-point functions, we note that the sign of the external momentum is important. 

%\subsection*{\underline{\bf{\it{Scalar Two-Point Function~:}}}}
\subsection*{Scalar Two-Point Function:}

\begin{equation}
\begin{split}
B_0(q_1^2;&m_1^2,m_2^2)
\\
&=\int\frac{d^4k}{(2\pi)^4}\frac{1}{[k^2-m_1^2+i\epsilon]}\frac{1}{[(k+q_1)^2-m_2^2+i\epsilon]}
\end{split}
\end{equation}

%\subsection*{\underline{Scalar Three-Point Function~:}}
\subsection*{Scalar Three-Point Function:}

\begin{equation}
\begin{split}
&C_0(q_1^2,q_2^2,q_3^2;m_1^2,m_2^2,m_3^2)
\\
&=\int\frac{d^4k}{(2\pi)^4}\frac{1}{[k^2-m_1^2+i\epsilon]}\frac{1}{[(k+q_1)^2-m_2^2+i\epsilon]}
\\
&\times\frac{1}{[(k+q_1+q_2)^2-m_3^2+i\epsilon]}
\end{split}
\end{equation}

%\subsection*{\underline{Scalar Four-Point Function~:}}
\subsection*{Scalar Four-Point Function:}

\begin{equation}
\begin{split}
&D_0(q_1^2,q_2^2,q_3^2,q_4^2,(q_1+q_2)^2,(q_1+q_4)^2,m_1^2,m_2^2,m_3^2,m_4^2)
\\
&=\int\frac{d^4k}{(2\pi)^4}\frac{1}{[k^2-m_1^2+i\epsilon]}\frac{1}{[(k+q_1)^2-m_2^2+i\epsilon]}
\\
&\times\frac{1}{[(k+q_1+q_2)^2-m_3^2+i\epsilon]}\frac{1}{[(k+q_1+q_2+q_3)^2-m_4^2+i\epsilon]}
\end{split}
\end{equation}

%\subsection*{\underline{Vector Three-Point Function~:}}
\subsection*{Vector Three-Point Function:}

\begin{equation}
\begin{split}
&C_1^\mu(q_1,q_2,q_3;m_1^2,m_2^2,m_3^2)
\\
&=\int\frac{d^4k}{(2\pi)^4}\frac{1}{[k^2-m_1^2+i\epsilon]}\frac{(2q_1+q_2+2k)^\mu}{[(k+q_1)^2-m_2^2+i\epsilon]}
\\
&\times\frac{1}{[(k+q_1+q_2)^2-m_3^2+i\epsilon]}
\end{split}
\end{equation}

%\subsection*{\underline{Vector Four-Point Function~:}}
\subsection*{Vector Four-Point Function:}

\begin{equation}
\begin{split}
&D_1^\mu(q_1,q_2,q_3,q_4,m_1^2,m_2^2,m_3^2,m_4^2)
\\
&=\int\frac{d^4k}{(2\pi)^4}\frac{1}{[k^2-m_1^2+i\epsilon]}\frac{(2q_1+q_2+2k)^\mu}{[(k+q_1)^2-m_2^2+i\epsilon]}
\\
&\times\frac{1}{[(k+q_1+q_2)^2-m_3^2+i\epsilon]}\frac{1}{[(k+q_1+q_2+q_3)^2-m_4^2+i\epsilon]}
\end{split}
\end{equation}

Making use of these definitions, we may write the loop correction expressions.

\section{Evaluation of Diagrams}\label{app:diagrams}

We begin with the two tree-level diagrams. The \\
$s$-channel diagram is:
\begin{equation}
i\mathcal{M}^{(a)\mu}=(-i\sqrt{2}g_{\pi N})S_F^N(p_s)(-ie)(p+p_s)^\mu.
\end{equation}
The $t$-channel diagram is:
\begin{equation}
i\mathcal{M}^{(g)\mu}=(-i\sqrt{2}g_{\pi N})S_F^\pi(p_t)(-ie)(p_t+p_\pi)^\mu .
\end{equation}
The two $s$-channel vertex corrections are:
\begin{equation}
\begin{split}
i\mathcal{M}^{(b)\mu}=&(-i\sqrt{2}g_{\pi N})S_F^N(p_s)
\\
&\times(-ie)\bigg[i g_{\pi N}^{\p2}C_1^\mu(p,q,-p_s;m_\pi^{\p2},m_N^{\p2},m_N^{\p2})\bigg],
\end{split}
\end{equation}
\begin{equation}
\begin{split}
i\mathcal{M}^{(c)\mu}=&(-i\sqrt{2}g_{\pi N})S_F^N(p_s)
\\
&\times(-ie)\bigg[2i g_{\pi N}^{\p2}C_1^\mu(p,q,-p_s;m_N^{\p2}m_\pi^{\p2},m_\pi^{\p2})\bigg].
\end{split}
\end{equation}
There are two self energy diagrams, corresponding to a virtual charged and neutral pion respectively:
\begin{equation}
\begin{split}
i\mathcal{M}^{(d)\mu}=&(-i\sqrt{2}g_{\pi N})S_F^N(p_s)\bigg[2g_{\pi N}^{\p2}B_0(p_s^2,m_N^{\p2},m_\pi^{\p2})\bigg]
\\
&\times S_F^N(p_s)(-ie)(p+p_s)^\mu ,
\end{split}
\end{equation}
\begin{equation}
\begin{split}
i\mathcal{M}^{(e)\mu}=&(-i\sqrt{2}g_{\pi N})S_F^N(p_s)\bigg[g_{\pi N}^{\p2}B_0(p_s^2,m_N^{\p2},m_\pi^{\p2})\bigg]
\\
&\times S_F^N(p_s)(-ie)(p+p_s)^\mu .
\end{split}
\end{equation}
The strong vertex is also modified by the loop corrections:
\begin{equation}
\begin{split}
i\mathcal{M}^{(f)\mu}=&(-i\sqrt{2}g_{\pi N})\bigg[ig_{\pi N}^{\p2}C_0(p_s^2,p_\pi^2,p^{\p2},m_\pi^{\p2},m_N^{\p2},m_N^{\p2})\bigg]
\\
&\times S_F^N(p_s)(-ie)(p+p_s)^\mu.
\end{split}
\end{equation}
There is one diagram which modifies the pion electromagnetic vertex:
\begin{equation}
\begin{split}
i\mathcal{M}^{(h)\mu}=&(-i\sqrt{2}g_{\pi N})S_F^\pi(p_t)
\\
&\times(-ie)\bigg[2ig_{\pi N}^{\p2}C_1^\mu(p_t,q,-p_\pi,m_N^{\p2},m_N^{\p2},m_N^{\p2})\bigg] ,
\end{split}
\end{equation}
and one self energy diagram:
\begin{equation}
\begin{split}
i\mathcal{M}^{(i)\mu}=&(-i\sqrt{2}g_{\pi N})S_F^\pi(p_t)\bigg[2g_{\pi N}^{\p2}B_0(p_t^2,m_N^{\p2},m_N^{\p2})\bigg]
\\
&\times S_F^\pi(p_t)(-ie)(p_t+p_\pi)^\mu .
\end{split}
\end{equation}
The $t$-channel strong vertex is also modified:
\begin{equation}
\begin{split}
i\mathcal{M}^{(j)\mu}=&(-i\sqrt{2}g_{\pi N})\bigg[ig_{\pi N}^{\p2}C_0(p^2,p_t^2,p^{\p2},m_\pi^{\p2},m_N^{\p2},m_N^{\p2})\bigg]
\\
&\times S_F^\pi(p_t)(-ie)(p_t+p_\pi)^\mu .
\end{split}
\end{equation}
At tree-level, there are no $u$-channel diagrams, since the neutron is neutral. However, quantum corrections modify the tree-level result. There are two corrections to the electromagnetic vertex:
\begin{equation}
\begin{split}
i\mathcal{M}^{(k)\mu}=&(-ie)\bigg[2ig_{\pi N}^{\p2}C_1^\mu(p_u,q,-p^\p;m_\pi^{\p2},m_N^{\p2},m_N^{\p2})\bigg]
\\
&\times S_F^N(p_u)(-i\sqrt{2}g_{\pi N}),
\end{split}
\end{equation}
\begin{equation}
\begin{split}
i\mathcal{M}^{(l)\mu}=&(+ie)\bigg[2ig_{\pi N}^{\p2}C_1^\mu(p_u,q,-p^\p;m_N^{\p2},m_\pi^{\p2},m_\pi^{\p2})\bigg]
\\
&\times S_F^N(p_u)(-i\sqrt{2}g_{\pi N}).
\end{split}
\end{equation}
Note that these come in with the opposite sign. Since we choose $m_\pi^\p=0.7$ GeV and $m_N^\p=0.71$ GeV, these two terms have approximately the same magnitude, but opposite sign. Thus the neutron's effect on the cross section is negligible. The single box diagram is:
\begin{equation}
\begin{split}
i\mathcal{M}^{(m)\mu}=&(-i\sqrt{2}g_{\pi N})(-ie)
\\
&\times\bigg[-g_{\pi N}^{\p2}D_1^\mu(p,q,-p_\pi,-p^\p,m_\pi^{\p2},m_N^{\p2},m_N^{\p2},m_N^{\p2})\bigg].
\end{split}
\end{equation}
%
%
%\pagebreak

\section{Fitted Cross Sections}\label{app:crossSections}

The $F_\pi$ Collaboration reported extracted pion form factor values for eight kinematic points. In our first analysis, we extracted the same kinematic points in our model. This required us to fit out model cross section to the pseudodata calculated using the loop-corrected cross section for each of the eight kinematic sets (see Table~\ref{table:kin_points}). The following eight plots show the agreement between the pseudodata, and our model cross section. In each case, the extracted pion form factor shown in Fig.~\ref{fig:extractedFormFactor} is obtained by evaluating the monopole form factor using the best fit for the monopole mass $\Lambda_\pi^2$.

We also include the calculated `strong' ($\pi N$) form factor which one encounters in the $t$-channel process. 
\begin{figure}[h]
\includegraphics[scale=0.4]{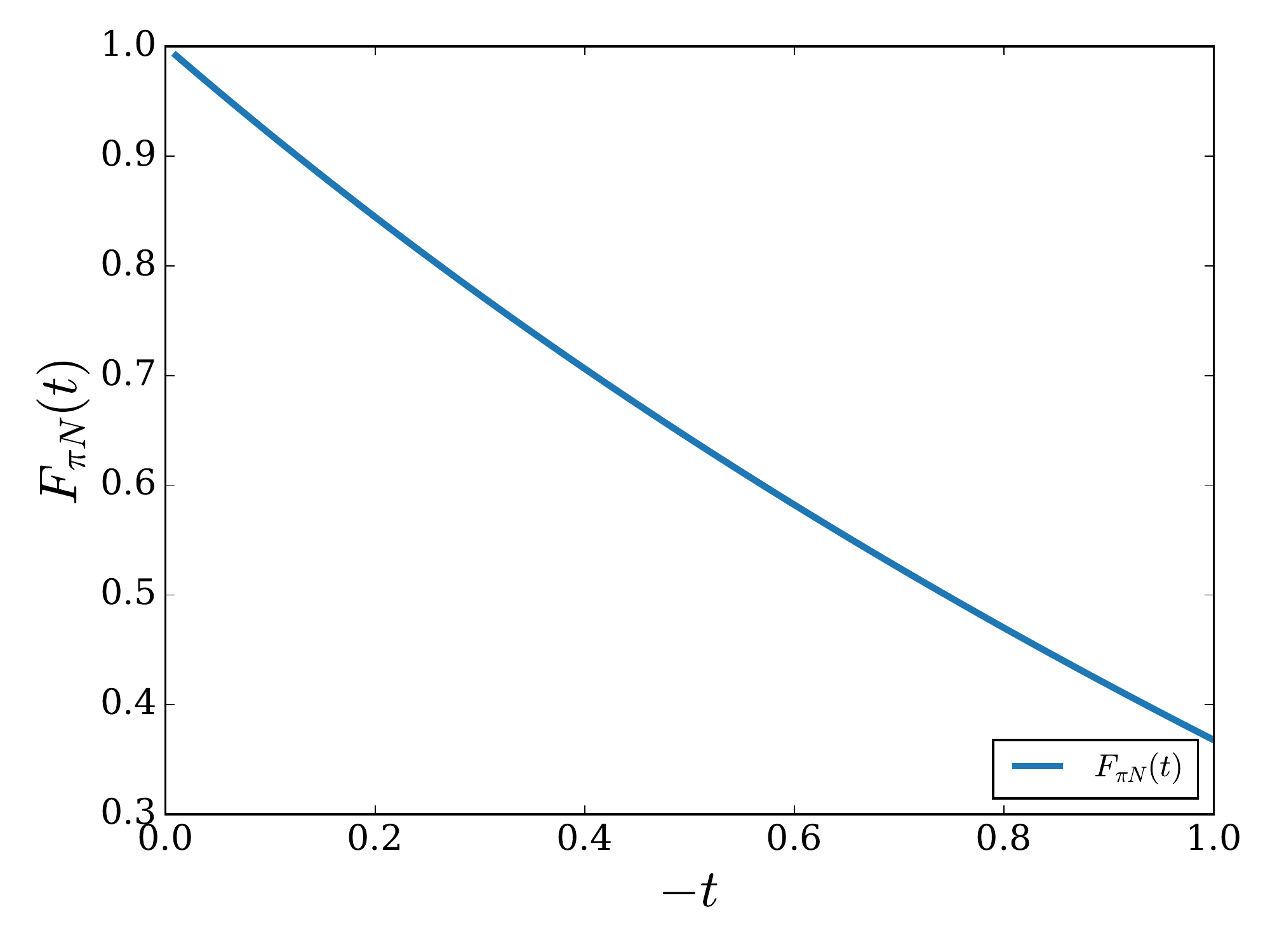}
\caption{(Colour online) Strong form factor predicted in toy model. Note that since we evaluate the cross section for small $|t|$, the main source of $t$ dependence in the cross section comes from the $t$-channel propagator.}
\end{figure}

\begin{figure*}[h]
\centering
\begin{subfigure}{0.49\textwidth}
\centering
\includegraphics[scale=0.40]{graphics/plots/Qs_035W_210_2}
\end{subfigure}
\begin{subfigure}{0.49\textwidth}
\centering
\includegraphics[scale=0.40]{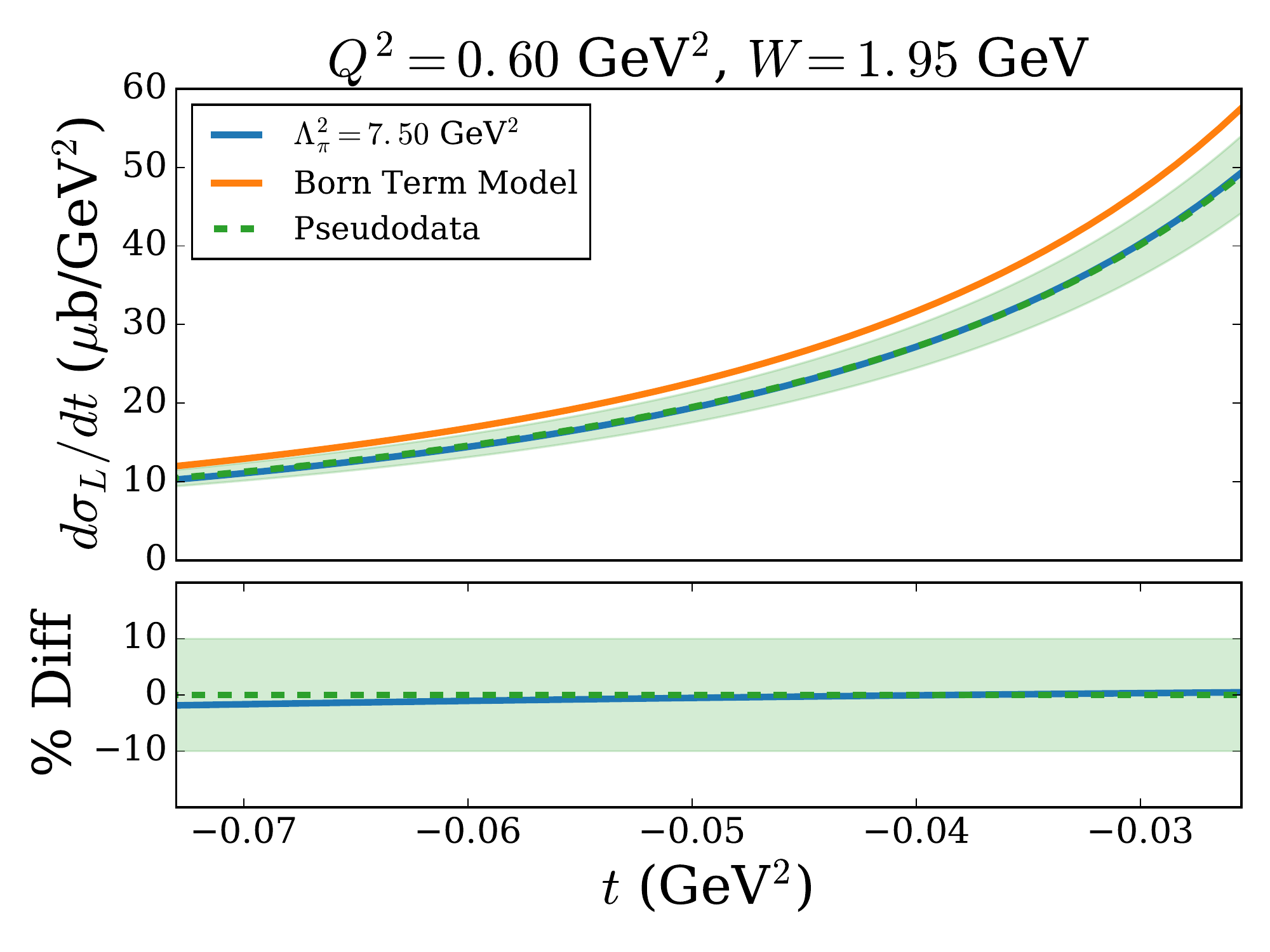}
\end{subfigure}
\vspace{-8pt}

\begin{subfigure}{0.49\textwidth}
\centering
\includegraphics[scale=0.40]{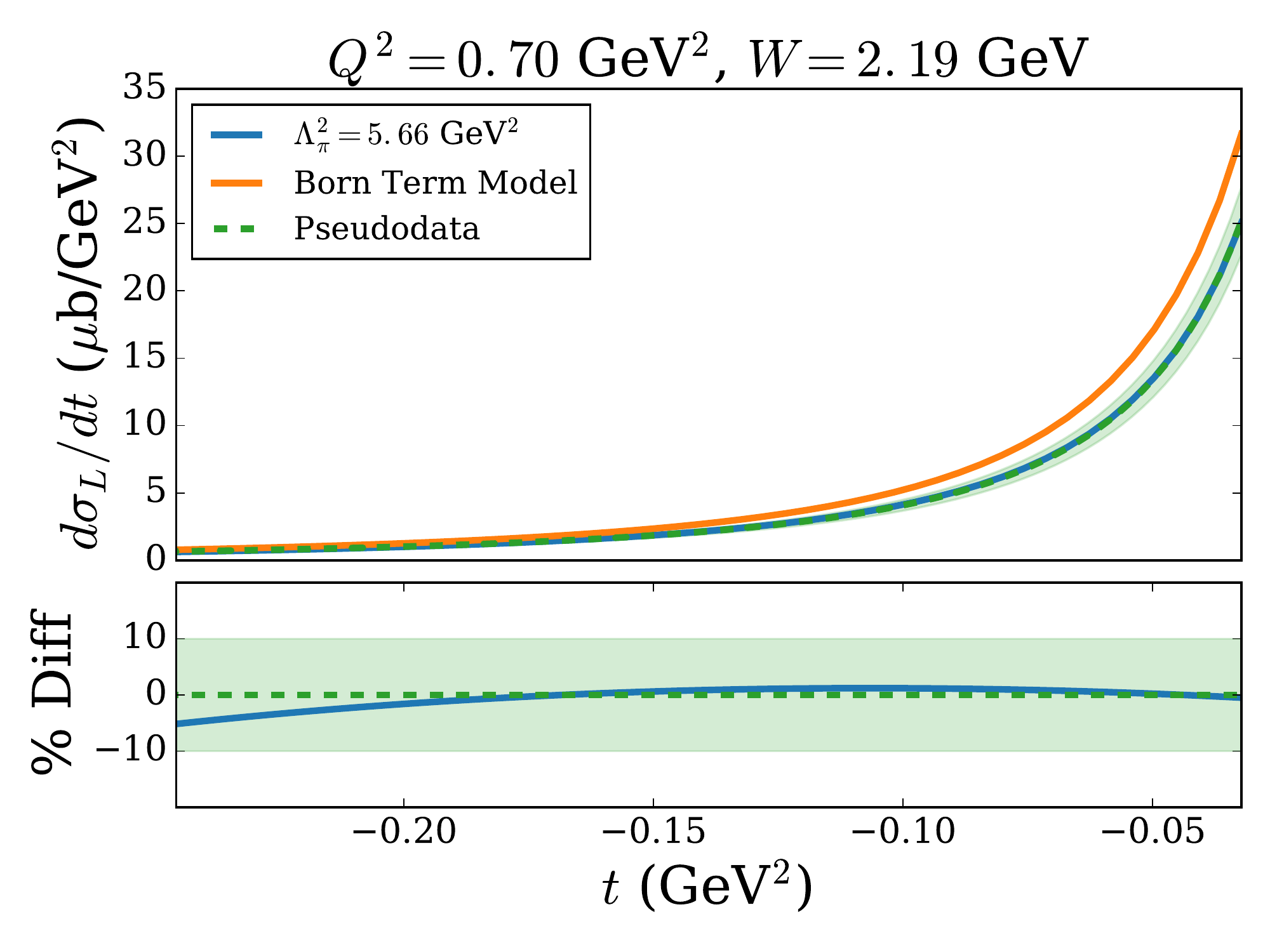}
\end{subfigure}
\begin{subfigure}{0.49\textwidth}
\centering
\includegraphics[scale=0.40]{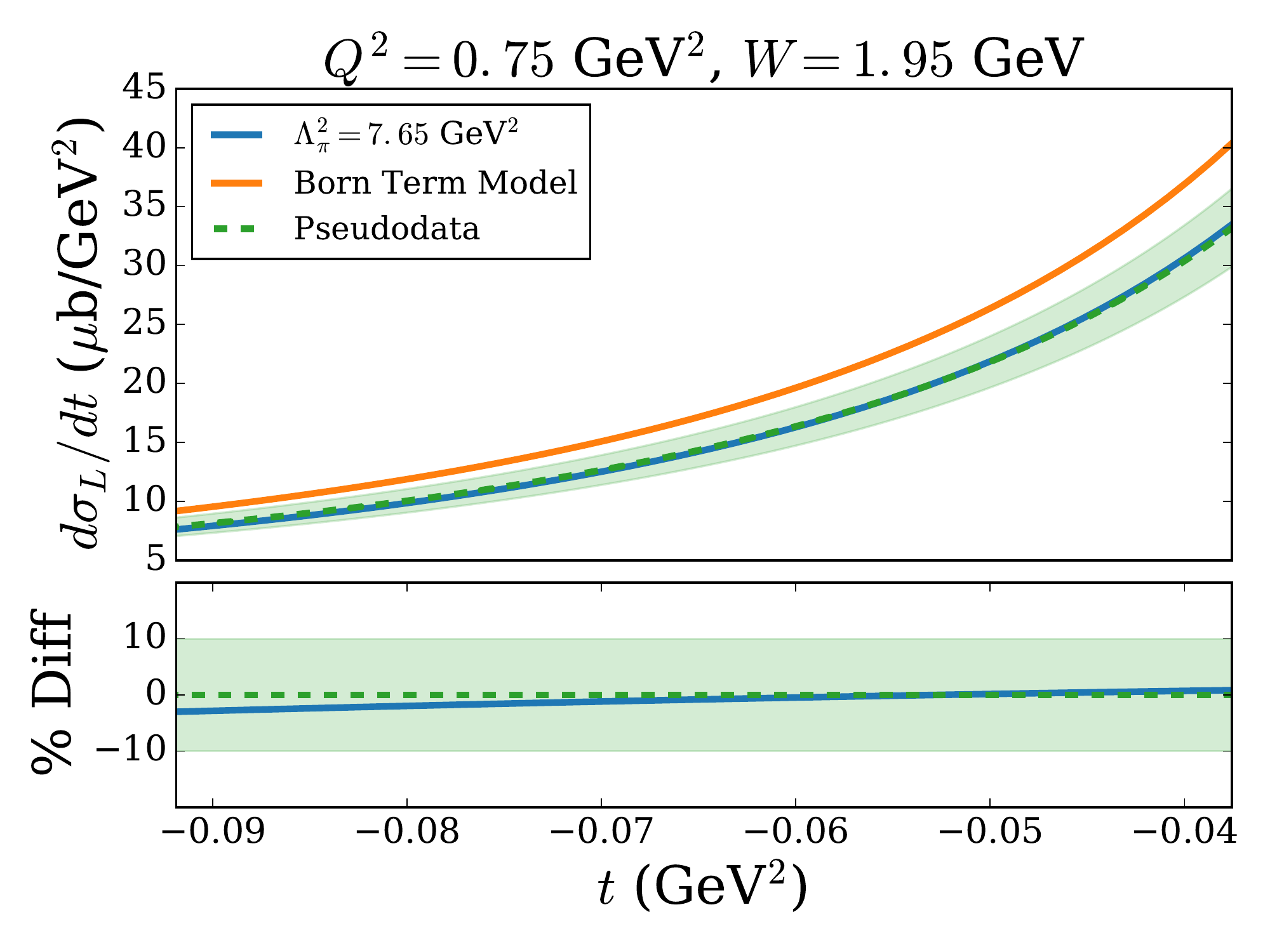}
\end{subfigure}
\vspace{-8pt}

\begin{subfigure}{0.49\textwidth}
\centering
\includegraphics[scale=0.40]{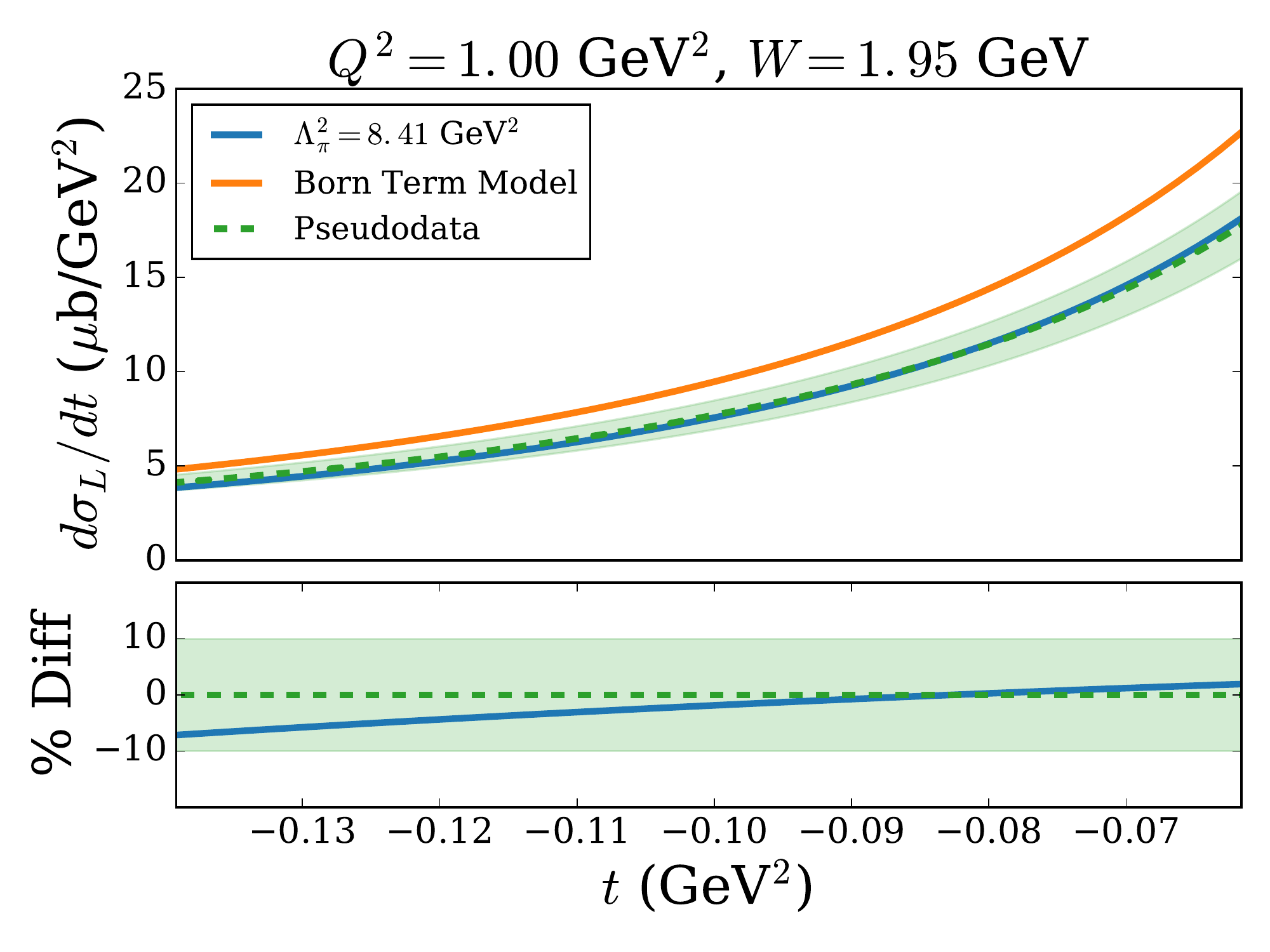}
\end{subfigure}
\begin{subfigure}{0.49\textwidth}
\centering
\includegraphics[scale=0.40]{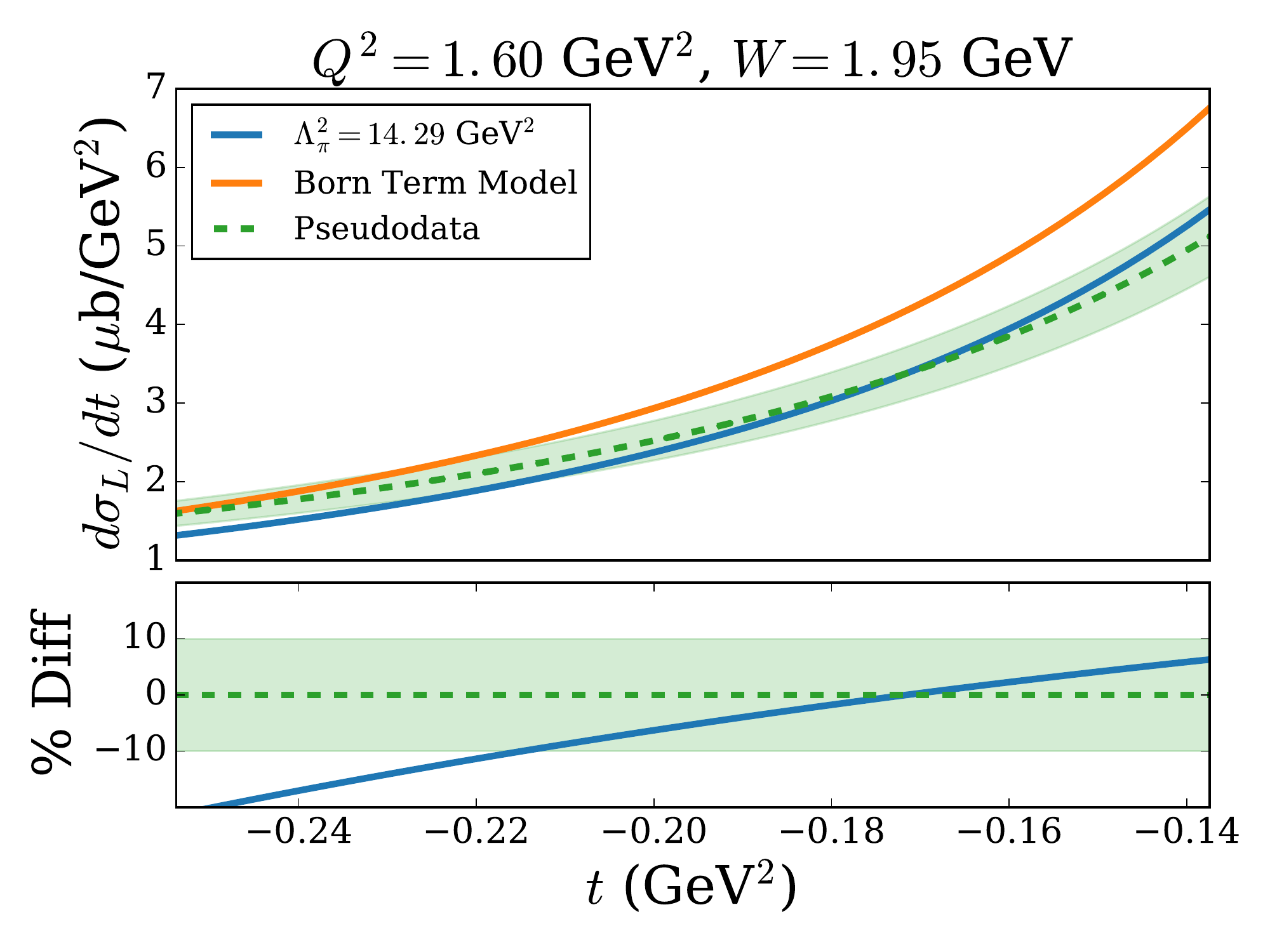}
\end{subfigure}
\vspace{-8pt}

\begin{subfigure}{0.49\textwidth}
\centering
\includegraphics[scale=0.40]{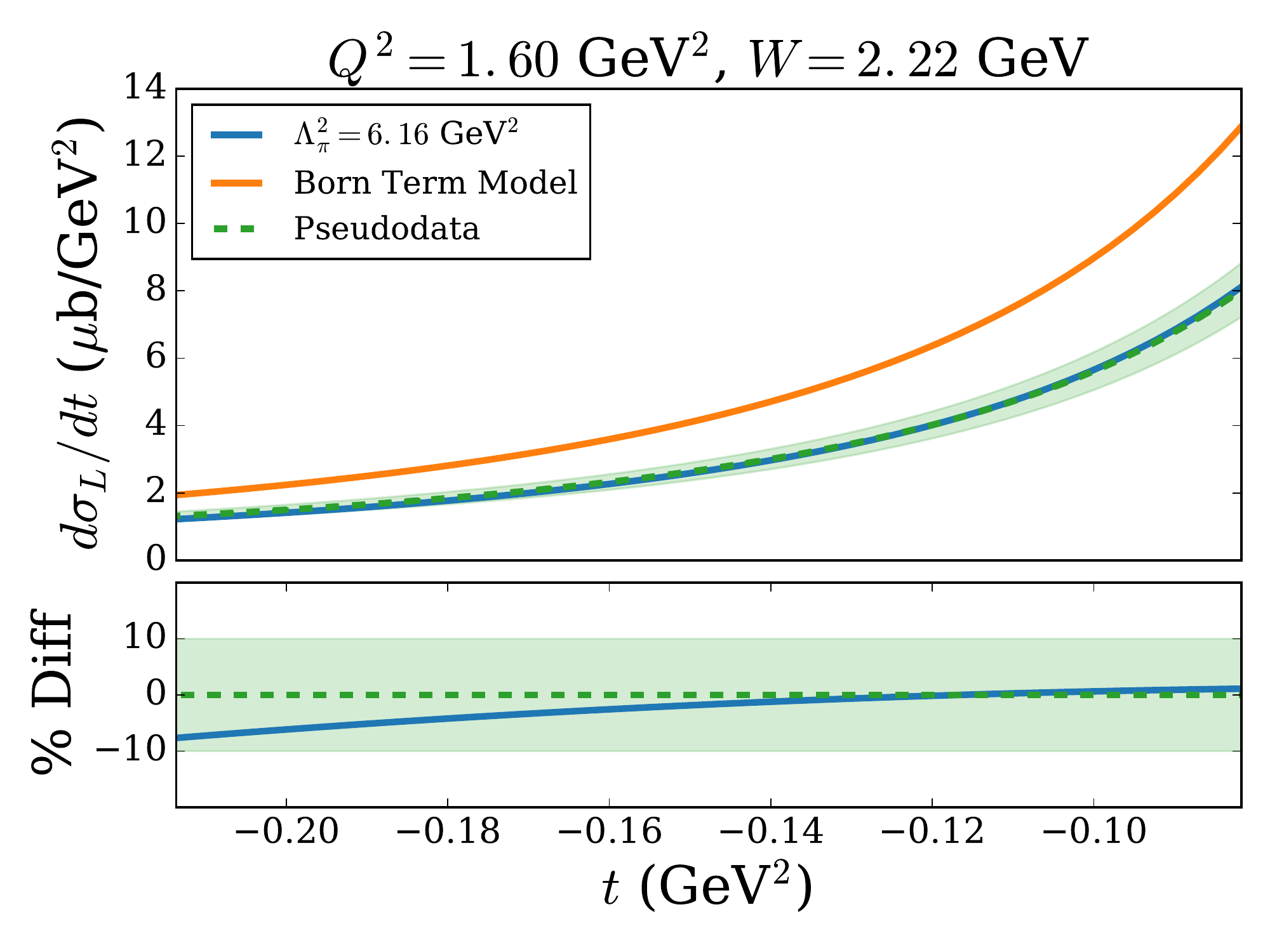}
\end{subfigure}
\begin{subfigure}{0.49\textwidth}
\centering
\includegraphics[scale=0.40]{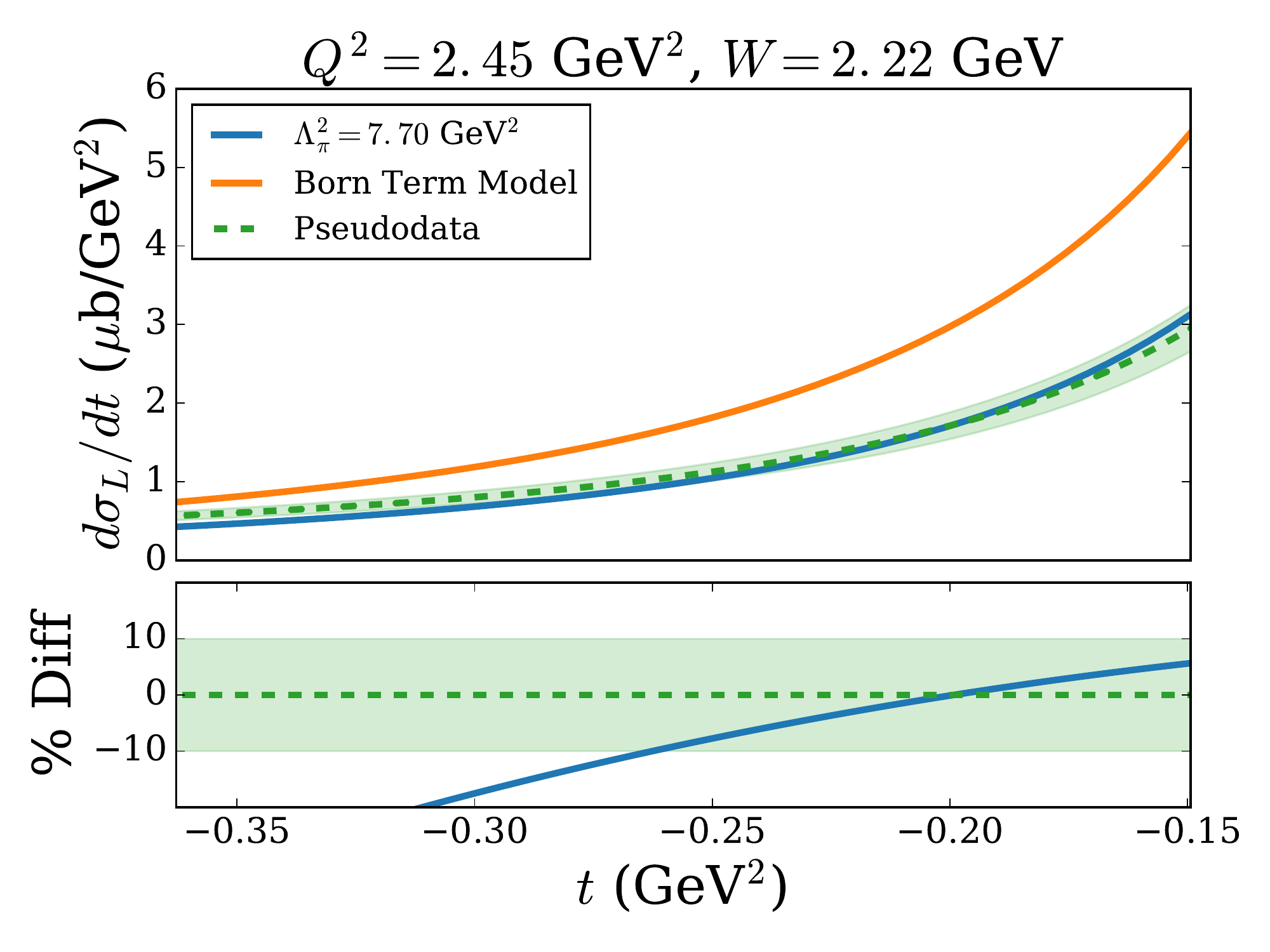}
\end{subfigure}
\vspace{-3mm}
\caption{(Colour online) Comparison of fitted model cross section to pseudodata.}
\label{fig:3}
\end{figure*}

\section*{References}

\bibliographystyle{model1-num-names}
\bibliography{Bibliography.bib}
%\bibliography{pionElectroproduction.bbl}

\end{fmffile}
\end{document}